\documentclass[11pt]{article}%

\usepackage{hyperref,cleveref, bm, commath}
\usepackage{graphicx,tabularx}
\usepackage{subcaption}
\usepackage[margin=1in]{geometry}

\usepackage[numbers]{natbib}

\usepackage{multirow}
\usepackage{algorithmic}

\usepackage{booktabs,siunitx}

\usepackage[ruled]{algorithm2e}
\SetKwInput{KwParam}{Parameter}
\SetAlgoCaptionLayout{centerline}
\usepackage{amsmath, amsfonts, amsthm}

\usepackage{xr}

\usepackage{lscape}

\usepackage{xcolor}

\usepackage{xr}
\usepackage{sectsty}
\usepackage[onehalfspacing]{setspace}
\newtheorem{proposition}{Proposition}

\newtheorem{assumption}{Assumption}
\crefname{assumption}{Assumption}{Assumptions}
\crefname{equation}{equation}{equations}

\theoremstyle{definition}

\makeatletter
\renewcommand{\algocf@captiontext}[2]{#1\algocf@typo. \AlCapFnt{}#2} 
\def\@algocf@capt@plain{top}
\renewcommand{\algocf@makecaption}[2]{%
  \addtolength{\hsize}{\algomargin}%
  \sbox\@tempboxa{\algocf@captiontext{#1}{#2}}%
  \ifdim\wd\@tempboxa >\hsize
  \hskip .5\algomargin%
  \parbox[t]{\hsize}{\algocf@captiontext{#1}{#2}}
  \else%
  \global\@minipagefalse%
  \hbox to\hsize{\box\@tempboxa}
  \fi%
  \addtolength{\hsize}{-\algomargin}%
}
\makeatother

\begin{document}
%
\sectionfont{\bfseries\large\sffamily}%
%

\subsectionfont{\bfseries\sffamily\normalsize}%
%

\noindent
{\sffamily\bfseries\Large
Powerful genome-wide design and robust statistical inference in two-sample summary-data
Mendelian randomization}%
%

\noindent
\textsf{Qingyuan Zhao\footnote{\textit{Address for correspondence:} Department of Statistics, The
Wharton School, University of Pennsylvania, Jon M. Huntsman Hall, 3730 Walnut
Street, Philadelphia, PA 19104-6340 USA. \ \textsf{E-mail:}
qyzhao@wharton.upenn.edu. \ Nov.\ 16, 2018.}, Yang Chen, Jingshu Wang, Dylan S. Small}%
%

\noindent
\textsf{University of Pennsylvania, University of Michigan, University
  of Pennsylvania, University of Pennsylvania}%

\noindent
\textsf{{\bf Abstract}: \noindent {\bf Background}.\ Mendelian randomization (MR) uses genetic
variants as instrumental variables to estimate
the causal effect of risk exposures in epidemiology. Two-sample
summary-data MR that uses publicly available genome-wide association
studies (GWAS) summary data have become a popular design in
practice. With the sample size of GWAS continuing to increase, it is now
possible to utilize genetic instruments that are only weakly associated
with the exposure.

\noindent {\bf Methods}.\ To maximize the statistical power of MR, we
propose a genome-wide design where more than a thousand genetic instruments are used. For the statistical
analysis, we use an empirical partially Bayes approach where
instruments are weighted according to their true strength, thus weak
instruments bring less variation to the estimator. The final estimator
is highly efficient in the presence of many weak genetic instruments and
is robust to balanced and/or sparse pleiotropy.

\noindent{\bf Results}.\ We apply our method to estimate the causal effect of
body mass index (BMI) and major blood lipids on cardiovascular disease
outcomes. Compared to previous MR studies, we obtain much more precise
causal effect estimates and substantially shorter confidence
intervals. Some new and statistically significant findings are: the
estimated causal odds ratio of BMI on ischemic stroke is 1.19 (95\%
CI: 1.07--1.32, $p$-value $\le 0.001$); the estimated causal odds ratio of
high-density lipoprotein cholesterol (HDL-C) on coronary artery
disease (CAD) is 0.78 (95\% CI 0.73--0.84, $p$-value $\le 0.001$). However,
the estimated effect of HDL-C becomes substantially smaller and statistically
non-significant when we only use the strong instruments.

\noindent{\bf Conclusions}.\ By employing a genome-wide design and
robust statistical methods, the statistical power of MR studies can be
greatly improved. Our empirical results suggest that, even though
the relationship between HDL-C and CAD appears to be highly
heterogeneous, it may be too soon to completely
dismiss the HDL hypothesis. Further investigations are needed to
demystify the observational and genetic associations between HDL-C and
CAD.

}%


\noindent
\textsf{{\bf Key messages}: \begin{itemize}
\item We utilize common variants across the whole genome (typically
  over a thousand) as instrumental variables.
\item We extend a previously proposed method---robust adjusted profile
  score---to account for the measurement error in GWAS summary data
  and biases due to weak instruments. A new method---empirical
  partially Bayes---is developed to increase the statistical power
  when some genetic instruments are strong but many are very weak. The
  estimator is robust to balanced and/or sparse pleiotropy.
\item Our new and more powerful analysis greatly improves the
  precision of the causal effect estimates of BMI and blood lipids on
  cardiovascular disease outcomes.
\item Code to replicate the results (including diagnostics) is available in the
  \texttt{R} package \texttt{mr.raps}
  (\url{https://github.com/qingyuanzhao/mr.raps}).
\end{itemize}
}

\newpage
\section{Introduction}
\label{sec:introduction}

Mendelian randomization (MR) is a method of using genetic variation to
infer the causal effect of a modifiable risk exposure on disease
outcome. Since MR can give unbiased estimates in the presence of unmeasured
confounding, it has become a widely used tool for epidemiologists and
health scientists \cite{davey2014mendelian}. A prominent example is
the overwhelming evidence of a causal link between low-density lipoprotein cholesterol
(LDL-C) and coronary heart disease found by several MR studies
\cite{linsel2008lifelong,ference2012effect,myocardial2014inactivating,burgess2016mendelian},
which is consistent with the results of earlier landmark clinical trials
\cite{scandinavian1994randomised}.


From a statistical perspective, MR is a special instance of
instrumental variable (IV) methods
\cite{hernan2006instruments,didelez2007mendelian}. Compared to
classical applications of the IV methods in economics
\cite{angrist1991does} and health research \cite{baiocchi2014instrumental},
the most distinctive feature of MR is the enormous number of candidate
instruments. Potentially, any one of the millions of single nucleotide
polymorphisms (SNP) in the human genome can be used as an IV as long
as it satisfies the following three validity criteria \cite{baiocchi2014instrumental}:
\begin{enumerate}
\item Relevance: the SNP must be associated with the risk
  exposure.
\item Effective random assignment: the SNP must be
  independent of any unmeasured confounder that is a common cause of
  the exposure and outcome under investigation.
\item Exclusion restriction (ER): the SNP affects outcome only
  through the risk exposure.
\end{enumerate}
Among the three criteria above, the ER assumption is most disputable
for MR due to a widespread phenomenon called pleiotropy
\cite{solovieff2013pleiotropy,verbanck2018detection},
a.k.a.\ multiple functions of genes. For example, there are many
SNPs that are associated with both LDL-C and high-density lipoproteins
cholesterol (HDL-C), thus their effects on
cardiovascular outcomes are possibly mediated by both lipids. When
these SNPs are used as instruments in a MR analysis of HDL-C, the ER
assumption is likely violated.

To alleviate these concerns, most existing MR
studies \cite{voight2012plasma,holmes2014mendelian,hemani2016mr}
select a handful of genome-wide significant SNPs
that are associated with the exposure risk factor and then seek to
justify that the ER assumption is reasonable. This simple design is
very transparent, but it has some major limitations. First, a full justification of the ER
assumption requires a deep understanding of the causal mechanism of
the genes and can be invalidated by new findings. For example,
Katan \cite{katan1986apoupoprotein}, an early exponent of MR, proposed to
use polymorphic forms of the \emph{APOE} gene
to estimate the causal effect of blood cholesterol on
cancer. However, as Davey Smith and Ebrahim \cite{davey2003mendelian}
later argued, they may be
invalid instruments due to pleiotropic effects on other biomarkers.
Second, the statistical power of MR is substantially reduced when the vast majority
of SNPs are
excluded, including a lot of known genetic variation
of the exposure and the outcome (\Cref{fig:dist-hdl}).

Meanwhile, it is well known to econometricians and statisticians
that weak IVs can still provide valuable information, especially if
there are a number of them
\cite{stock2002survey,hansen2008estimation}. 
In a previous paper \cite{zhao2018statistical},
we found that using
weakly significant SNPs can greatly increase the
efficiency of MR studies. In a different but related application,
Bulik-Sullivan et al.\ \cite{bulik2015atlas} also found that a genome-wide analysis is much
more powerful than using just the significant SNPs to estimate the
genetic correlation.
Using weak instruments also helps to test the presence of
effect heterogeneity \citep{bowden2018improving} and
identify candidate IVs that do not satisfy the ER
assumption, and the causal effect can still be consistently estimated
when the invalid IVs are rare or the pleiotropic effects are balanced
\cite{zhao2018statistical,kang2016instrumental,bowden2016consistent,guo2016confidence}.

In this paper, we will introduce two new strategies that can greatly
increase the statistical power of MR studies. The first innovation is a
truly genome-wide design: unlike previous MR studies including our
earlier work \cite{zhao2018statistical}, no
threshold will be used in the selection phase (apart from demanding the
genetic instruments to be independent). 
Typically, about 1000 independent genetic instruments will be used in
a genome-wide MR study. Notice that a genetic variant is considered to
satisfy the first IV assumption (relevance) even if it does not
causally modify the exposure. The variant can be used as an IV if it is in
linkage disequilibrium with a causal variant and thus associated with
the exposure \cite{hernan2006instruments}. The
blueprint of genome-wide MR has been discussed in the
literature before
\cite{evans2013mining,brion2014beyond,evans2015mendelian}, but it was
not feasible until recently because most existing summary-data MR
methods are heavily biased by weak IVs.

Our second innovation is an estimator that adaptively assigns
weights to the IVs according to their strength. 
This method is based on a general empirical
partially Bayes approach introduced by Lindsay
\cite{lindsay1985using}. Our previous
method of adjusting the profile score \cite{zhao2018statistical}
can be viewed as a special case of this approach using a predetermined
flat prior. Both approaches yield consistent and asymptotically
normal estimators of the causal effect, but using an empirically
estimated prior can substantially increase the statistical
power. The structure of the empirical partially Bayes approach
also motivates a new diagnostic plot and a hypothesis test that is
useful to detect effect heterogeneity according to instrument
strength.


\section{Genome-wide design}
\label{sec:datas-data-struct}

We will use a working example to illustrate
the genome-wide MR design, where the goal is to estimate the
causal effect of body mass index (BMI) on the risk of CAD. Increased
adiposity was found to increase the risk of CAD in several previous MR studies
\citep{holmes2014causal,hagg2015adiposity,dale2017causal,lyall2017association}. In
this paper we use this example as a positive control to demonstrate
how using weak instruments can greatly improve the precision of MR.

Our two-sample summary-data MR design makes use of three
non-overlapping GWAS:
\begin{enumerate}
\item \texttt{Selection dataset}: A GWAS for BMI in the Japanese
  population involving $173,430$ individuals \cite{akiyama2017genome};
\item \texttt{Exposure dataset}: A GWAS for BMI in the UK BioBank
  involving more than $350,000$ individuals \cite{nealelabukbb};
\item \texttt{Outcome dataset}: A GWAS for CAD conducted by the
  CARDIoGRAMplusC4D consortium of about $185,000$ cases and controls
  with genotype imputation using the 1000 Genomes Project
  \cite{nikpay2015comprehensive}.
\end{enumerate}
For each GWAS, the summary data are publicly available, which
report the linear or logistic regression coefficients and standard
errors (typically following a meta-analysis) of all the genotyped or
imputed SNPs. The \texttt{selection} and \texttt{exposure} datasets
are two non-overlapping GWAS for the same (or similar) phenotypes. We
recommend to reserve the GWAS with higher quality (e.g.\ larger sample size,
same population as the \texttt{outcome dataset}) for the
\texttt{exposure} dataset. The quality of the \texttt{selection
  dataset} is often less important with a genome-wide MR design.

After obtaining the GWAS summary datasets, we preprocess the data to
select genetic instruments for the statistical analysis. We first
remove SNPs that do not coappear in all three datasets. Then we use the remaining
\texttt{selection dataset} to find independent SNPs (distance $\ge$ 10 megabase
pairs, linkage disequilibrium $R^2 \le 0.001$) that are most associated
with BMI. This is done in a greedy fashion using the
linkage-disequilibrium (LD) clumping function in the PLINK software
\cite{purcell2007plink}. Using independent
SNPs makes the statistical analysis more convenient and is common in
MR studies \cite{hemani2016mr}. Suppose $p$ SNPs
are selected after LD clumping. Usually $p$ is about $1000$ after the
described preprocessing; in our working example concerning BMI and CAD, $p = 1119$.

A typical summary-data MR dataset thus consists of $2p$ marginal genetic effect estimates
(linear/logistic regression
coefficients) and their standard errors from the exposure and outcome
datasets:
\begin{itemize}
\item $\hat{\gamma}_j$, $j=1,\dotsc,p$ are the genetic effects on the
  exposure (BMI). The standard errors are denoted by
  $\sigma_{Xj}$.
\item $\hat{\Gamma}_j$, $j=1,\dotsc,p$ are the genetic effects on the
  outcome (CAD). The standard errors are denoted by
  $\sigma_{Yj}$.
\end{itemize}

The study design described above is usually called a two-sample summary-data
MR study \cite{burgess2015using}. Here we want to emphasize that using a separate and
non-overlapping dataset for SNP selection is very important for the
unbiasedness of the genetic effect estimates, thus eliminating any
bias due to ``winner's curse'' \cite{burgess2011avoiding}. A common
misconception is that, when the same GWAS is used for selection and
to obtain $\hat{\gamma}_j$, the ``winner's curse'' could be avoided
by only using genome-wide significant SNPs. This is not true, because
although these SNPs are most likely true hits, the estimated genetic
effects $\hat{\gamma}_j$ are still biased. As a consequence, the
causal effect estimate is generally biased towards zero \cite{zhao2018statistical}.

\begin{figure}
  \centering
  \includegraphics[width = 0.8\textwidth]{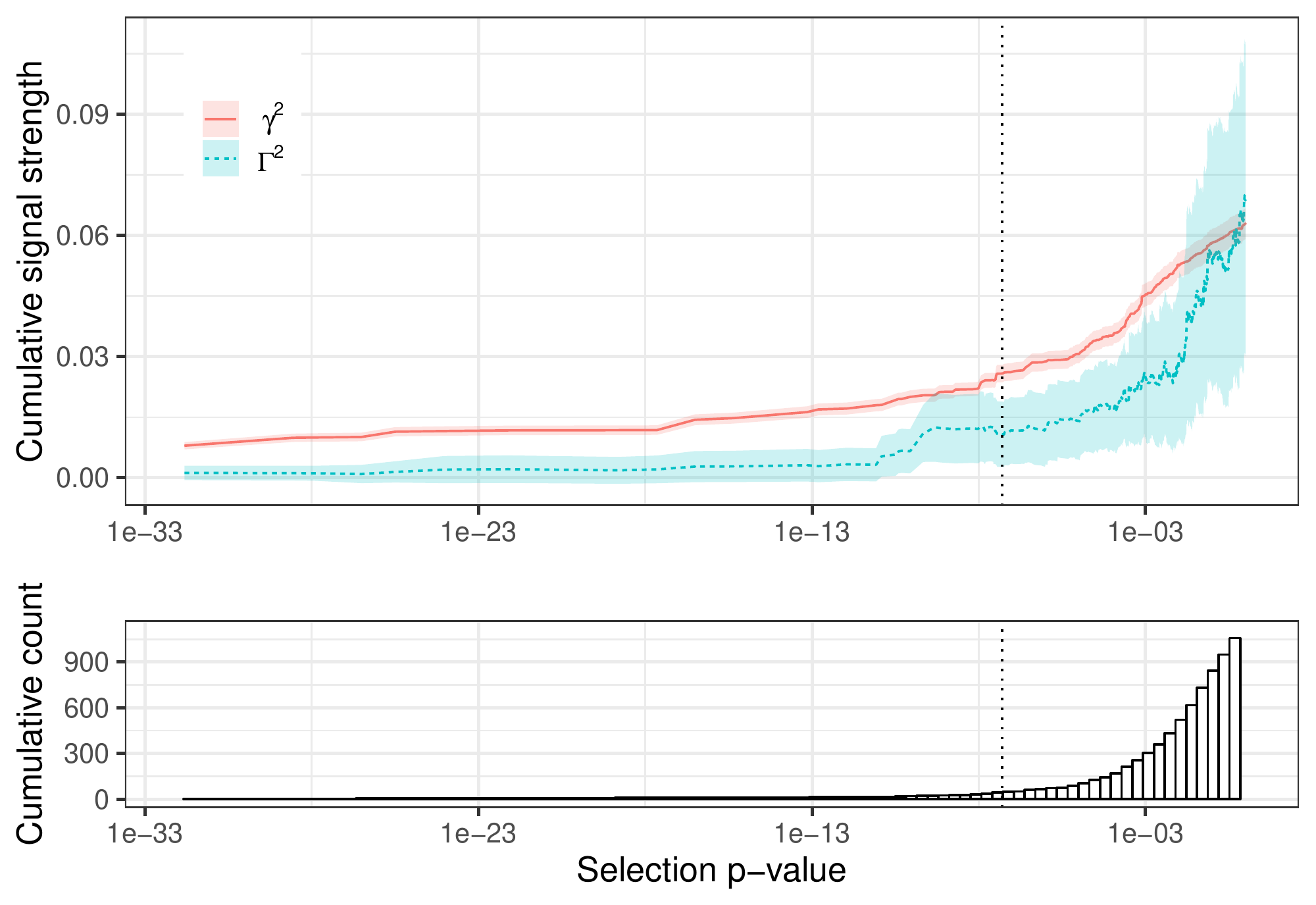}
  \caption{Distribution of signal strength in the genome-wide MR of
    Body Mass Index (BMI) on Coronary Artery Disease (CAD) as a function
    of the selection $p$-value: $\sum_{\text{SNP}~j~\text{pass
        threshold}} \gamma_j^2$ and $\sum_{\text{SNP}~j~\text{pass
        threshold}} \Gamma_j^2$. See \Cref{sec:datasets} below for more details of
    this dataset. Shaded region in the main plot is 95\% confidence
    interval of the estimated cumulative signal strength. The dotted
    vertical line corresponds to the standard genome-wide significance threshold
    $5 \times 10^{-8}$ in the \texttt{selection dataset}.}
  \label{fig:dist-hdl}
\end{figure}

\Cref{fig:dist-hdl} shows the distribution of genetic signal strength
measured by the squared $l_2$ norm $\|\bm \gamma\|^2 = \sum_{j=1}^p
\gamma_j^2$ (for BMI) and $\|\bm \Gamma\|^2$
(for CAD) as a function of the selection threshold. Throughout the
paper we use bold letters to indicate vectors. These quantities
are closely related to the genetically inherited phenotypic variance and can be unbiasedly estimated by $\sum_{j} \hat{\gamma}_j^2 - \sigma_{Xj}^2$
and $\sum_{j} \hat{\Gamma}_j^2 - \sigma_{Yj}^2$ over the SNPs that
pass the selection threshold. Compared to the
conventional analysis that only uses 44 genome-wide significant SNPs
($p$-value $\le 5 \times 10^{-8}$), the genome-wide MR design using
all the 1119 SNPs contains almost twice amount of genetic variation
for BMI. This observation suggests a great potential of
increasing the statistical power of the MR analysis by utilizing the
weaker instruments.

\section{Statistical model}
\label{sec:meas-error-model}

Following our previous article \cite{zhao2018statistical}, our main modeling assumptions are:
\begin{assumption}[Measurement error model] \label{assump:measurement-error}
  \[
  \begin{pmatrix}
    \hat{\bm \gamma} \\
    \hat{\bm \Gamma}
  \end{pmatrix}
  \sim \mathrm{N}\left(
    \begin{pmatrix}
      \bm \gamma \\
      \bm \Gamma
    \end{pmatrix},
    \begin{pmatrix}
      \bm{\Sigma}_X & \bm 0 \\
      \bm 0 & \bm{\Sigma}_Y \\
    \end{pmatrix}\right),~\bm \Sigma_X =
  \mathrm{diag}(\sigma_{X1}^2,\dotsc,\sigma_{Xp}^2),~\bm \Sigma_Y =
  \mathrm{diag}(\sigma_{Y1}^2,\dotsc,\sigma_{Yp}^2).
  \]
\end{assumption}
\begin{assumption}[Pleiotropy model] \label{assump:robust-mr}
  We assume the causal effect $\beta$ satisfies $\Gamma_j \approx
  \beta \gamma_j$ for most $j = 1,\dotsc,p$. More specifically, let $\bm \alpha = \bm \Gamma - \beta \bm \gamma$. We assume
  $\alpha_j$ is independent of $\gamma_j$ and most
  $\alpha_j \overset{\mathrm{ind}}{\sim} \mathrm{N}(0, \tau^2)$ where
  $\tau^2$ is a small overdispersion parameter. A small proportion of
  the SNPs (say 5\%) might
  deviate from this model and have very large $|\alpha_j|$.
\end{assumption}

The normality and independence assumptions in
\Cref{assump:measurement-error} can be immediately justified by the large
sample size of GWAS, non-overlapping samples in the \texttt{selection}, \texttt{exposure}, and
\texttt{outcome datasets}, and independence of the SNPs
\cite{zhao2018statistical}. We have also implicitly assumed that the
standard errors $\sigma_{Xj}$, $\sigma_{Yj}$, $j=1,\dotsc,p$ reported
in the GWAS are well calibrated.

\Cref{assump:robust-mr} presumes that for most SNPs, the genetic
associations with the exposure and the outcome approximately satisfy a
pair-wise linear relationship, with the common slope parameter being
the causal effect $\beta$. When all the SNPs are
valid IVs, the linear relationship $\bm \Gamma = \beta \bm
\gamma$ can be derived through assuming the SNPs, exposure and outcome
variables follow a linear structural model
\cite{bowden2017framework}. This result can be extended to nonlinear structural
models by assuming the per-SNP effects are minuscule and the SNPs affect
the exposure in a homogenous way \cite{zhao2018statistical}. When
the outcome variable is the binary disease status, $\beta$ may be
interpreted as a conservative estimate of the causal log-odds-ratio
\cite{zhao2018statistical}.

In reality, many genetic
variants may violate the ER assumption and have other pathways to
affect CAD. For example, a SNP selected for the MR study of HDL-C
might also be associated with LDL-C, so its genetic association with
CAD includes both the causal effect of HDL-C (if any) and
LDL-C. Motivated by the exploratory analysis for the
MR of adiposity on blood pressure in our previous paper
\cite{zhao2018statistical}, the robust MR
model in \Cref{assump:robust-mr} considers two types of deviations
from the exact linear relationship $\bm \Gamma = \beta \bm \gamma$:
1.\ small and balanced pleiotropy represented by the random effects
model $\alpha_j \sim \mathrm{N}(0,\tau^2)$; 2.\ idiosyncratic
and large pleiotropy. The first kind of deviation is a special
case of the InSIDE (Instrument Strength Independent of Direct Effect)
assumption \cite{bowden2017framework,bowden2015mendelian}, while the
second is similar to the sparse invalid IV
assumption \cite{kang2016instrumental}. We think it is crucial that
the statistical method of MR is robust to both kinds of pleiotropy.

As a remark, when the InSIDE assumption is not satisfied, the causal
effect $\beta$ cannot be identified without further assumptions. In this
case, the estimand of our statistical method below is $\beta$ plus
$(\bm \gamma^T \bm \gamma)^{-1} \bm \gamma^T \bm \alpha$, the
regression slope of $\bm \alpha$ on $\gamma$.

Before diving into the details of our statistical methodology, we want
to mention that an alternative approach to handle widespread
pleiotropy is the multivariable MR
\cite{burgess2015multivariable,sanderson2018examination}, where
several exposures are examined simultaneously. In the rest of this
paper we will focus on the genome-wide design for univariate MR and
explore the multivariate extension in a forthcoming work.

\section{Statistical method}
\label{sec:linds-empir-part}

In our previous article \cite{zhao2018statistical}, we proposed a
robust estimator based on adjusting the profile score function of $\beta$
and $\tau^2$, in which the nuisance parameters $\bm \gamma$ are
profiled out. Here we propose a new method to eliminate the nuisance
parameters $\bm \gamma$ based on an empirical partially Bayes approach
introduced by Lindsay \cite{lindsay1985using}. The new method has a
simple geometric interpretation and can further increase the
statistical power.

\subsection{Empirical partially Bayes}
\label{sec:empir-part-bayes}

We first consider the simplest scenario: $\bm \alpha = \bm 0$. The key
insight is gained from deriving the contribution of the
$j$-th SNP to the conditional score function (\Cref{sec:technical-details}):
\begin{equation*} \label{eq:conditional-score}
  C_j(\beta,\gamma_j) = \frac{\gamma_j(\hat{\Gamma}_j - \beta
    \hat{\gamma}_j)}{\beta^2 \sigma_{Xj}^2 + \sigma_{Yj}^2}.
\end{equation*}
It is straightforward to verify that the maximum likelihood estimator
(MLE) of $\gamma_j$ given $\beta$, denoted as
$\hat{\gamma}_{j,\mathrm{MLE}}(\beta)$, is a sufficient statistic of
$\gamma_j$ and is independent of $\hat{\Gamma}_j - \beta
\hat{\gamma}_j$. The decoupling of ``instrument strength''
$\gamma_j$ and ``regression residual'' $\hat{\Gamma}_j - \beta
\hat{\gamma}_j$ in $C_j(\beta,\gamma_j)$ motivates
us to consider a general class of estimating functions:
\[
C(\beta) = \sum_{j=1}^p \frac{f_j(\beta,\hat{\gamma}_{j,\mathrm{MLE}}(\beta))
  \cdot \psi(t_j(\beta))}{\sqrt{\beta^2 \sigma_{Xj}^2 +
    \sigma_{Yj}^2}},~t_j(\beta) = \frac{\hat{\Gamma}_j - \beta
  \hat{\gamma}_j}{\sqrt{\beta^2 \sigma_{Xj}^2 +
  \sigma_{Yj}^2}},
\]
where $f_j$ is an arbitrary function of $\beta$ and
$\hat{\gamma}_{j,\mathrm{MLE}}$, and $\psi$ is an odd function (so
$\psi(-t) = -\psi(t)$). Because
$\hat{\gamma}_{j,\mathrm{MLE}}(\beta)$ is independent of $\hat{\Gamma}_j - \beta
\hat{\gamma}_j$ (and thus $t_j(\beta)$), it is easy to show that
$\mathbb{E}[C(\beta)] = 0$ at the true $\beta$. Therefore the root of
$C(\beta)$, denoted by $\hat{\beta}$, is a reasonable estimator of
$\beta$. Geometrically, this estimating function finds the $\beta$
such that a transformation (by $f$) of the estimated ``instrument strength''
$\hat{\gamma}_{j,\mathrm{MLE}}(\beta)$ is uncorrelated with a
transformation (by $\psi$) of the ``regression residual'' $\hat{\Gamma}_j - \beta
\hat{\gamma}_j$.

Different choices of the weighting scheme $f_j$ do not change the
unbiasedness of $C(\beta)$, but may affect the statistical
efficiency. The profile score developed in our previous article
\cite{zhao2018statistical} amounts to using
$\hat{\gamma}_{j,\mathrm{MLE}}(\beta)$ as the weight. To maximize
statistical power, Lindsay
\cite{lindsay1985using} suggested to use the
empirical Bayes estimate of $\gamma_j$ as the weight,
\[
f_j(\beta,\hat{\gamma}_{j,\mathrm{MLE}}(\beta)) =
\hat{\gamma}_{j,\mathrm{EB}}(\beta) =
\mathbb{E}_{\pi_{\hat{\eta}}} \big[\gamma_j|\hat{\gamma}_{j,\mathrm{MLE}}(\beta)\big]
\]
where $\pi_{\eta}$ is a prior distribution of $\bm \gamma$ and
$\hat{\eta}$ is an empirical estimate of the prior
parameter. Intuitively, $\hat{\gamma}_{j,\mathrm{EB}}$ shrinks
$\hat{\gamma}_{j,\mathrm{MLE}}$ towards $0$. The function $\psi$ is
chosen to limit the influence of large outliers
(\Cref{sec:robust-estimator}).

\subsection{Spike-and-slab prior}
\label{sec:spike-slab-prior}

In principle, a good choice of the prior distribution $\pi_{\eta}$
should have the following properties: 1.\ the parametric family
$\pi_{\eta}$ should fit the distribution of $\bm \gamma$ reasonably
well, so we can gain efficiency by using the empirical partially Bayes
estimator; 2.\ The empirical Bayes estimator of $\gamma_j$ should be easy to
compute since it will be evaluated many times when iteratively
solving the estimating equations. For these reasons, we choose to use a
spike-and-slab Gaussian mixture prior \cite{mitchell1988bayesian,george1993variable} to model
$\gamma_j/\sigma_{Xj}$ in all our empirical examples:
\begin{equation*} \label{eq:spike-slab}
  \gamma_j/\sigma_{Xj} \sim \pi_{p_1,\sigma_1,\sigma_2} = p_1 \cdot \mathrm{N}(0,
  \sigma_1^2) + (1 - p_1) \cdot \mathrm{N}(0, \sigma_2^2).
\end{equation*}
We decide to model the effect sizes $\gamma_j/\sigma_{Xj}$ instead of
the effects $\gamma_j$ because this scale is more familiar and the
shrinkage rule is easier to interpret. Typically, $p_1$ is close to
$1$, $\sigma_1^2$ is close to zero (the spike component), and
$\sigma_2^2$ is much larger than $\sigma_1^2$ (the slab
component). The selective shrinkage offered by the spike-and-slab
prior \cite{ishwaran2005spike} is essential to gain efficiency in
empirical partially Bayes (\Cref{sec:impl-deta}).

\subsection{Robust estimator}
\label{sec:robust-estimator}

To account for invalid IVs in \Cref{assump:robust-mr}, we need to further estimate the overdispersion parameter
$\tau^2 = \mathrm{Var}(\alpha_j)$ while being robust to large outliers
of $\alpha_j$. Intuitively, we need two estimating equations after
eliminating the nuisance $\bm \gamma$: one for $\beta$ and one for
$\tau^2$. For $\beta$, we can follow the empirical partially Bayes
approach described above by replacing $\sigma_{Yj}^2$ with
$\sigma_{Yj}^2 + \tau^2$. For $\tau^2$,
we need to adjust the profile score function of $\tau^2$ due to a
Neyman-Scott phenomenon
\cite{zhao2018statistical,neyman1948consistent}. To be robust against
outliers, we propose to use a bounded function of the
``regression residual'' $\hat{\Gamma}_j - \beta \hat{\gamma}_j$.

Next we describe the robust estimating function of $\beta$ and
$\tau^2$. Derivation of these functions is very similar to our
previous RAPS (Robust Adjusted Profile Score) approach
\cite{zhao2018statistical} and the details are omitted. Let
$\psi_1(\cdot)$ and $\psi_2(\cdot)$ be two differentiable odd
functions. The empirical
partially Bayes version of the RAPS estimator
$(\hat{\beta},\hat{\tau}^2)$ is given by the solution
to $\tilde{\bm C}(\beta,\tau^2) = (\tilde{C}_1(\beta,\tau^2),
\tilde{C}_2(\beta,\tau^2))^T = \bm 0$, where
\begin{equation*} \label{eq:eb-estimating-function}
  \begin{split}
    \tilde{C}_1(\beta,\tau^2) &=
    \sum_{j=1}^p\frac{\hat{\gamma}_{j,\mathrm{EB}}(\beta,\tau^2) \cdot
      \psi_1\big(t_j(\beta,\tau^2)\big)}{s_j(\beta,\tau^2)},~\text{and} \\
    \tilde{C}_2(\beta,\tau^2) &=
    \sum_{j=1}^p\frac{\psi_2\big(t_j(\beta,\tau^2)\big) -
      \delta}{s_j^2(\beta,\tau^2)},~\text{where}~\delta = \mathbb{E}[\psi_2(Z)]~\text{for}~Z \sim
    \mathrm{N}(0,1),
  \end{split}
\end{equation*}
where $
t_j(\beta,\tau^2) = (\hat{\Gamma}_j - \beta
\hat{\gamma}_j)/s_j(\beta,\tau^2)$ is the standardized regression residual
and $s_j(\beta,\tau^2) = \sqrt{\beta^2 \sigma_{Xj}^2 + \sigma_{Yj}^2 + \tau^2}$.

In the empirical examples below, we will use $\psi_2(t) = t \cdot \psi_1(t)$
and consider two choices of $\psi_1$: the non-robust identity
function $\psi_I$ and the robust Huber's score function $\psi_H$. A
situation we sometimes encounter with real data is that the RAPS
estimating equation may have several roots. In this case, we report
the root closest to the optimization-based profile-likelihood
estimator if there no other close root; otherwise we simply report the
empirical partially Bayes estimator is not available. We do not
interpret this as a defect of the proposed method; rather, there is
often strong evidence for effect heterogeneity in this situation and we
think the investigator should avoid making any hasty
conclusion. Further implementation details including how to compute
the standard error of $\hat{\beta}$ can be found in \Cref{sec:impl-deta}.

\subsection{Diagnostics}
\label{sec:diagnostics}

A potential concern of using many weak instruments is that they might have
more pleiotropy than strong instruments
\cite{verbanck2018detection,jordan2018landscape}. It is possible that
the covariance $\mathrm{Cov}(\alpha_j,\gamma_j)$ is a function of the
instrument strength $\gamma_j$ and is non-zero for certain range of
$\gamma_j$ (under \Cref{assump:robust-mr} or more generally the InSIDE
assumption, $\mathrm{Cov}(\alpha_j,\gamma_j) \approx 0$). In general,
the conclusions of a genome-wide MR study are stronger if the weak instruments and strong
instruments produce similar estimates of the causal effect. We propose
a simple diagnostic plot for this purpose, where the standardized
regression residual $t_j(\beta,\tau^2)$ is plotted against (a
standardized version of) the estimated instrument strength
\[
\hat{\gamma}_{j,\mathrm{EB}}(\beta,\tau^2)\big/\sqrt{\mathrm{Var}\big(\hat{\gamma}_{j,\mathrm{EB}}(\beta,\tau^2)\big)},
\]
evaluated at $(\beta,\tau^2) = (\hat{\beta}, \hat{\tau}^2)$. At the
true value of $(\beta,\tau^2)$ and when our modeling assumptions are
satisfied, $t_j$ should be independent of
$\hat{\gamma}_{j,\mathrm{EB}}$ and follow a standard normal
distribution (possibly with some outliers). This proposition can be
empirically checked in a diagnostic plot
(\Cref{sec:diagnostics,sec:diagnostic-plots}). Furthermore, it can
used to test for the heterogeneity of the instruments. In our
empirical studies, we report the heterogeneity $p$-value as the null
test for the linear regression of standardized residuals on estimated
instrument strength (expanded using B-splines with degrees of freedom
equal to $p/20$). This can be used as a falsification test of our
modeling assumptions. A related but different graphical outlier detection method is the
radial MR plot proposed by Bowden et al.\ (2018) \cite{bowden2017improving2}.

\section{Datasets}
\label{sec:datasets}

\Cref{tab:gwas-datasets} summarizes the datasets used in our empirical studies. We
will consider 6 phenotypes: BMI, LDL-C, HDL-C, triglycerides (TG), CAD
(or myocardial infarction), and ischemic stroke (IS), and apart from
IS, all other phenotypes have at least two non-overlapping GWAS
results. \Cref{tab:list-mr-analyses} lists the design of the empirical
examples in the next two sections.

\def\arraystretch{1.2}

\begin{table}[htbp]
  \centering \small
  \begin{subtable}[t]{\textwidth}\centering
    \begin{tabular}{ccccc}
      \toprule
      Phenotype & Dataset name & Citation or data source & Population &
      Sample size \\
      \midrule
      \multirow{2}{*}{BMI} & \texttt{BMI (Jap)} & Akiyama et al.\ (2017) \cite{akiyama2017genome}
      & Asian & 173,430 \\
      \cline{2-5}
      & \texttt{BMI (UKB)} & UK BioBank \cite{nealelabukbb} & European
      & $>350,000$ \\
      \midrule
      \multirow{6}{*}{Lipids} & \texttt{LDL} (2010) &
      \multirow{3}{*}{Teslovich et al.\ (2010) \cite{teslovich2010biological}} & \multirow{3}{*}{European}
      & \multirow{3}{*}{$\sim 100,000$} \\
      & \texttt{HDL (2010)} & & & \\
      & \texttt{TG (2010)} & & & \\
      \cline{2-5}
      & \begin{tabular}{c} \texttt{LDL (2013)} \\ \texttt{HDL (2013)} \\ \texttt{TG (2013)} \end{tabular} &
      \begin{tabular}{c} Global Lipids Genetics \\ Consortium (2013)
        \cite{willer2013discovery} \end{tabular} & European
      & $\sim 100,000$ \\
      \midrule
      \multirow{5}{*}{CAD} & \texttt{CARDIoGRAM} & Schunkert et al.\ (2011)
      \cite{schunkert2011large} & European & 86,995 \\
      \cline{2-5}
      & \texttt{C4D} & \begin{tabular}{c} C4D Genetics \\ Consortium (2011)
        \cite{coronary2011genome} \end{tabular} & \begin{tabular}{c} European and \\
        South Asian \end{tabular} & 30,442 \\
      \cline{2-5}
      & \texttt{CAD} & \begin{tabular}{c} CARDIoGRAMplusC4D \\ Consortium (2015)
        \cite{nikpay2015comprehensive} \end{tabular} & European & $\sim
      185,000$ \\
      \midrule
      \begin{tabular}{c} Myocardial\\ Infarction\end{tabular} & \texttt{MI (UKB)}
      & UK BioBank \cite{nealelabukbb} & European
      & $>350,000$ \\
      \midrule
      \begin{tabular}{c} Ischemic \\ Stroke \end{tabular} & \texttt{IS} & Malik et al.\ (2018) \cite{malik2018multiancestry} &
      European & 446,696 \\
      \bottomrule
    \end{tabular}
    \caption{List of publicly available GWAS summary datasets used in this
      paper. See \Cref{sec:data-avail} for web links we used to download the
      datasets.}
    \label{tab:gwas-datasets}
  \end{subtable}

  \begin{subtable}[t]{\textwidth}\centering \vspace{10pt}
    \begin{tabular}[h]{ccccc}
      \toprule
      Study name & \begin{tabular}{c} Screening \\ dataset \end{tabular} & \begin{tabular}{c} Exposure \\ dataset \end{tabular} & \begin{tabular}{c} Outcome \\ dataset \end{tabular} &
      Results \\
      \midrule
      Simulations & \multicolumn{3}{c}{\begin{tabular}{c} Data are
          simulated to mimic the MR \\ analysis for the LDL-CAD study
          below.\end{tabular}} & \Cref{sec:simulations}; \Cref{tab:sim} \\
      \midrule
      CAD-CAD & \texttt{MI (UKB)} & \texttt{C4D} &
      \texttt{CARDIoGRAM} & \Cref{sec:cad-cad-study}; \Cref{tab:cad-cad} \\
      \midrule
      BMI-CAD & \multirow{2}{*}{\texttt{BMI (Jap)}} & \multirow{2}{*}{\texttt{BMI (UKB)}} & \texttt{CAD} &
      \multirow{2}{*}{\begin{tabular}{c} \Cref{sec:bmi-results}; \Cref{tab:bmi-results,tab:bmi-detailed};\\ \Cref{fig:ije-results,fig:bmi-diagnostics}\end{tabular}} \\
      BMI-IS & & & \texttt{IS} & \\
      \midrule
      LDL-CAD & \texttt{LDL (2010)} & \texttt{LDL (2013)} & \multirow{3}{*}{\texttt{CAD}}
      & \multirow{3}{*}{\begin{tabular}{c} \Cref{sec:results};
          \Cref{tab:primary-summary,tab:lipid1-detailed,tab:lipid2-detailed};\\
          \Cref{fig:ije-results,fig:lipid-diagnostics-1,fig:lipid-diagnostics-2}
        \end{tabular}} \\
      HDL-CAD & \texttt{HDL (2010)} & \texttt{HDL (2013)} & &
      \\
      TG-CAD & \texttt{TG (2010)} & \texttt{TG (2013)} & & \\
      \midrule
      LDL-MI & \texttt{LDL (2010)} & \texttt{LDL (2013)} &
      \multirow{3}{*}{\texttt{MI (UKB)}}
      & \multirow{3}{*}{\begin{tabular}{c} \Cref{sec:results};
          \Cref{tab:primary-summary,tab:lipid3-detailed,tab:lipid4-detailed};\\
          \Cref{fig:lipid-diagnostics-3,fig:lipid-diagnostics-4}
        \end{tabular}} \\
      HDL-MI & \texttt{HDL (2010)} & \texttt{HDL (2013)} & &
      \\
      TG-MI & \texttt{TG (2010)} & \texttt{TG (2013)} & & \\
      \bottomrule
    \end{tabular}
    \caption{List of MR analyses.}
    \label{tab:list-mr-analyses}
  \end{subtable}
  \caption{GWAS Datasets and MR analyses in this paper.}
  \label{tab:list}
\end{table}

In our genome-wide MR design, we require the \texttt{selection},
\texttt{exposure}, and \texttt{outcome} GWAS to have non-overlapping
samples. Often the GWAS summary results are based on meta analyses of
smaller cohorts, so this assumption can be examined by checking if
they have no common participating cohort. This is the case for most of
our empirical examples below. The only exception is the MR analysis of
the blood lipids and CAD, where the \texttt{outcome dataset} obtained
from the CARDIoGRAMplusC4D consortium overlaps with the lipids
GWAS reported by the GLGC consortium. Nevertheless, the sample overlap
appears to be small by examining the participating cohorts.

\begin{figure}[t]
  \centering
  \includegraphics[width = \textwidth]{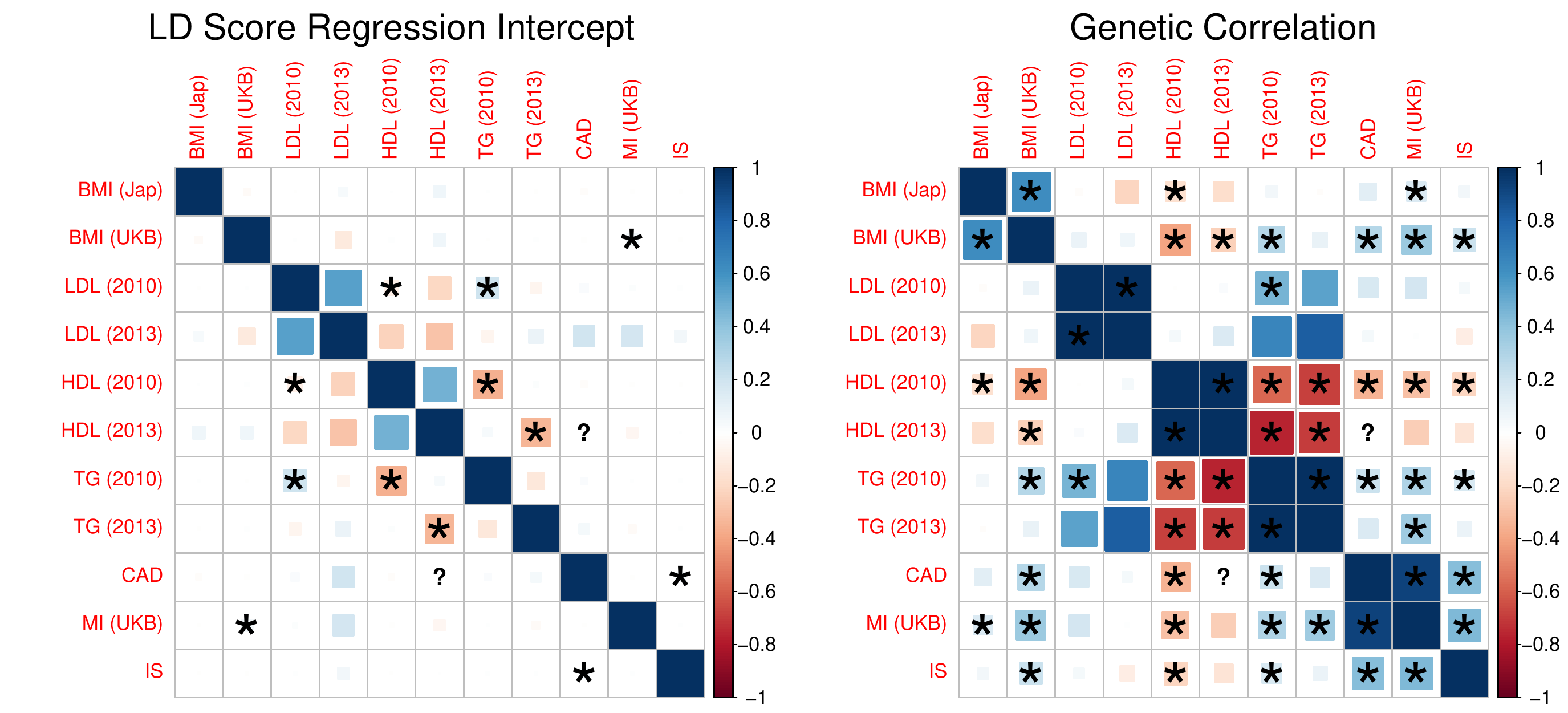}
  \caption{Results of LD score regression for the GWAS
    datasets. Significant intercepts and slopes (adjusted for multiple
    testing) are indicated by asterisks.}
  \label{fig:ld-score}
\end{figure}

The amount of sample overlap can also be tested by running the LD score
regression \cite{bulik2015ld}. Bulik-Sullivan et al.\ (2015)
\cite{bulik2015atlas} show that the intercept in the LD score
regression is proportional to the amount of sample overlap, and the
slope in the regression is proportional to the genetic
correlation. \Cref{fig:ld-score} shows the results of the LD score
regression for the datasets used in this paper. The regression
intercepts between the lipids datasets and CAD are relatively small
and not statistically significant, indicating the amount of sample
overlap may be small.

\section{Validation studies}
\label{sec:validation-studies}

\subsection{Simulations}
\label{sec:simulations}

We perform two validation studies of the proposed statistical
method. The first is a simulation study that mimics the real data
analysis for LDL-C and CAD using restricted instruments
(\Cref{tab:lipid3-ldl}). 
The GWAS summary data are simulated according to
\Cref{assump:measurement-error,assump:robust-mr} in six settings: NOO (No
Outlier), ALL (All SNPs), STR
(Strong SNPs), WKS (Weak SNPs), NUL (null causal effect), and EXP
(exponetially distributed $\gamma$). The true causal effect $\beta$ is
set to $0.2$ in all settings beside NUL (where $\beta = 0$). This is smaller
than the estimated causal effect of LDL-C so that the statistical
power is not always $100\%$. Except for EXP, the effect sizes
$\gamma_j/\sigma_{Xj},~j=1,\dotsc,p$ are generated from the Gaussian
mixture distribution with
\begin{description}
\item[Settings NOO, ALL, NUL] $p = 898$, $p_1 = 0.92$, $\sigma_1 = 0.47$, $\sigma_2 = 3.48$,
  resembling the analysis of LDL-C using all the $898$ SNPs.
\item[Setting STR] $p = 11$, $p_1 = 0.01$, $\sigma_1 = \sigma_2 =
  5.93$, resembling the analysis using the $11$ SNPs that are
  genome-wide significant in the \texttt{selection dataset}.
\item[Setting WKS] $p=887$, $p_1 = 0.92$, $\sigma_1 = 0.44$, $\sigma_2
  = 2.39$, resembling the analysis using the $887$ SNPs that are not
  significant in the \texttt{selection dataset}.
\end{description}
In setting EXP, the effect sizes are generated from the Laplace
distribution with rate $1.5$ (i.e.\ mean absolute value is $2/3$).

Next, $\bm{\Gamma}$ is
generated by $\Gamma_j = \beta \gamma_j + \alpha_j$ where $\alpha_j$
is an independent Gaussian variable with mean $0$ and variance $3.8
\times 10^{-5}$, the estimated $\tau^2$ in the LDL-CAD study using all
$898$ SNPs. In settings ALL, STR, and WKS, we include an outlier (substract
$5\tau$ from the $\alpha_j$ corresponding to the strongest SNP) to test
the method's robustness to large idiosyncratic pleiotropy. Finally,
the standard deviations $(\sigma_{Xj},\sigma_{Yj})$ are the same as
the standard errors in the analysis of LDL-C.

\def\arraystretch{1.2}

\begin{table}[htbp] \centering \setlength{\tabcolsep}{6pt} \small
  \centering
  \begin{tabular}{ll|ccccc}
    \toprule
    \multirow{3}{*}{\begin{tabular}{c} Setting \\ (\# SNPs, Outlier) \end{tabular}} & \multirow{3}{*}{Metric} &
    \multicolumn{5}{c}{Method} \\
    \cline{3-7}
    & & \multirow{2}{*}{MR-Egger} & MR-Egger & \multirow{2}{*}{Wtd.\ Med.} & RAPS & RAPS \\
    & & & (SIMEX) & & (MLE) & (Shrinkage) \\
    \midrule
    NOO ($898$ SNPs, 0 outlier) & Mean  & 0.13 & 0.15 & 0.12 & 0.20 & 0.20 \\
    & RMSE  & 0.088 & 0.073 & 0.095 & 0.073 & 0.063 \\
    & Coverage  & 61.1\% & 80.0\% & 74.2\% & 94.4\% & 95.3\% \\
    & Power  & 78.6\% & 86.2\% & 56.3\% & 80.2\% & 88.9\% \\
    \cline{3-7}
    ALL ($898$ SNPs, 1 outlier) & Mean  & 0.11 & 0.13 & 0.10 & 0.18 & 0.17 \\
    & RMSE  & 0.105 & 0.089 & 0.117 & 0.082 & 0.073 \\
    & Coverage  & 47.4\% & 65.8\% & 60.2\% & 91.8\% & 90.7\% \\
    & Power  & 64.8\% & 74.9\% & 40.2\% & 71.1\% & 79.1\% \\
    \cline{3-7}
    STR ($11$ SNPs, 1 outlier) & Mean  & 0.08 & 0.09 & 0.11 & 0.12 & 0.12 \\
    & RMSE  & 0.237 & 0.163 & 0.167 & 0.144 & 0.145 \\
    & Coverage  & 85.9\% & 81.1\% & 81.4\% & 84.9\% & 84.7\% \\
    & Power  & 12.2\% & 20.5\% & 27.3\% & 26.0\% & 26.0\% \\
    \cline{3-7}
    WKS ($887$ SNPs, 1 outlier) & Mean  & 0.08 & 0.10 & 0.07 & 0.18 & 0.17 \\
    & RMSE  & 0.134 & 0.116 & 0.144 & 0.119 & 0.106 \\
    & Coverage  & 38.3\% & 57.6\% & 50.6\% & 94.1\% & 92.6\% \\
    & Power  & 33.6\% & 46.4\% & 19.1\% & 39.4\% & 44.9\% \\
    \cline{3-7}
    NUL ($898$ SNPs, 0 outlier) & Mean  & 0.00 & 0.00 & 0.00 & 0.00 & -0.00 \\
    & RMSE  & 0.045 & 0.053 & 0.054 & 0.074 & 0.065 \\
    & Coverage  & 93.7\% & 92.5\% & 95.9\% & 93.6\% & 94.1\% \\
    & Power  & 6.3\% & 7.5\% & 4.1\% & 6.4\% & 5.9\% \\
    \cline{3-7}
    EXP ($898$ SNPs, 0 outlier) & Mean  & 0.10 & 0.11 & 0.08 & 0.20 & 0.20 \\
    & RMSE  & 0.112 & 0.111 & 0.135 & 0.085 & 0.082 \\
    & Coverage  & 50.7\% & 65.7\% & 45.8\% & 94.3\% & 95.2\% \\
    & Power  & 54.7\% & 50.7\% & 22.4\% & 69.9\% & 70.5\% \\
    \bottomrule
  \end{tabular}
  \caption{Simulation results using 1000 replications in 6
    settings: NOO
    (no outlier), ALL (all SNPs), STR (strong SNPs), WKS (weak SNPs),
    NUL (null causal effect), EXP (exponentially
    distributed effect sizes). For each estimator we
    report four metrics: mean of
    $\hat{\beta}$ (the true $\beta = 0$ in scenario NUL and $\beta =
    0.2$ in all other cases), root-mean-squared error
    (RMSE), coverage of the 95\% CI, and statistical power (proportion
    of 95\% CI not covering 0).}
  \label{tab:sim}
\end{table}

\setlength{\tabcolsep}{2pt}
\begin{table}[ht] \small \centering
  \begin{tabular}{lcccccc}
    \hline
    & \multicolumn{2}{c}{$p_\mathrm{sel} \in (0, 1)$} & \multicolumn{2}{c}{$p_\mathrm{sel} \in (0, 5 \times 10^{-8})$} & \multicolumn{2}{c}{$p_\mathrm{sel} \in (5 \times 10^{-8}, 1)$} \\
    \hline
    \# SNPs  & \multicolumn{2}{c}{1650} & \multicolumn{2}{c}{5} & \multicolumn{2}{c}{1645} \\
    $p_1$  & \multicolumn{2}{c}{0.99} & \multicolumn{2}{c}{0.01} & \multicolumn{2}{c}{0.88} \\
    $\sigma_1$  & \multicolumn{2}{c}{0.44} & \multicolumn{2}{c}{1.58} & \multicolumn{2}{c}{0.25} \\
    $\sigma_2$  & \multicolumn{2}{c}{4.5} & \multicolumn{2}{c}{6.96} & \multicolumn{2}{c}{1.25} \\
    MR-Egger  & \multicolumn{2}{c}{0.353 (0.033)} &
    \multicolumn{2}{c}{0.744 (0.476)} & \multicolumn{2}{c}{0.274
      (0.034)} \\
    MR-Egger (SIMEX)  & \multicolumn{2}{c}{0.744 (0.054)} & \multicolumn{2}{c}{0.764 (0.559)} & \multicolumn{2}{c}{0.592 (0.057)} \\
    Wtd.\ Med.  & \multicolumn{2}{c}{0.127 (0.035)} & \multicolumn{2}{c}{0.664 (0.125)} & \multicolumn{2}{c}{0.089 (0.034)} \\
    \hline 
    & MLE & Shrinkage & MLE & Shrinkage & MLE & \multicolumn{1}{c}{Shrinkage} \\
    $\tau^2 = 0$, $\psi_I$  & 1.029 (0.081) & 1.042 (0.076) & 0.937 (0.098) & 0.936 (0.098) & 1.058 (0.106) & 1.147 (0.104) \\
    $\tau^2 = 0$, $\psi_H$  & 1.058 (0.085) & 1.097 (0.08) & 0.822 (0.096) & 0.821 (0.096) & 1.076 (0.112) & 1.178 (0.109) \\
    $\tau^2 \ne 0$, $\psi_I$  & 1.029 (0.081) & 1.042 (0.076) & 0.952 (0.178) & 0.952 (0.178) & 1.058 (0.106) & 1.147 (0.104) \\
    $\tau^2 \ne 0$, $\psi_H$  & 1.055 (0.133) & 1.096 (0.107) & 0.926 (0.193) & 0.926 (0.193) & 1.076 (0.112) & 1.178 (0.109) \\
    \hline
  \end{tabular}
  \caption{Validation CAD-CAD study: both the exposure and the outcome are
    coronary artery disease, so the true $\beta$ should be about
    $1$. Our estimators are roughly unbiased and the shrinkage
    estimator is about 10\% more efficient when all SNPs are used.}
  \label{tab:cad-cad}
\end{table}

\Cref{tab:sim} shows the results of this simulation study for five
estimators: MR-Egger \cite{bowden2015mendelian}, MR-Egger with SIMEX
correction \citep{bowden2016assessing}, weighted median
\cite{bowden2016consistent}, RAPS with MLE weights, and RAPS with
shrinkage weights (both RAPS estimator use the Huber score function as $\psi$). In
settings with no outlier (NOO, NUL, EXP), the RAPS
estimators are unbiased and have the correct confidence interval
coverage. This shows that our empirical partially Bayes approach
remains unbiased even if the effect size distributions are
misspecified. The SIMEX correction helps to reduce the weak instrument
bias of MR-Egger but does not completely eliminate it. In other
settings with a large outlier (ALL, STR, WKS), the two RAPS
estimators have much smaller bias and are generally much more precise
than the other methods. Among the two RAPS estimator, the one with
shrinkage weights is about 7.5\% more efficient in setting EXP and
about 25\% more efficient in settings NOO, ALL and WKS where the ``spike''
and the ``slab'' have greater disparity.

\subsection{CAD-CAD study}
\label{sec:cad-cad-study}

Our second validation study uses a design in our previous article
\cite{zhao2018statistical} where the ``causal effect'' $\beta$ can be
regarded as known. In this example, all three
datasets---\texttt{selection}, \texttt{exposure}, \texttt{outcome} (as explained in
\Cref{sec:datas-data-struct})---are GWAS of coronary artery disease. In
particular, \texttt{MI (UKB)} is used to select SNPs, while
\texttt{C4D} and \texttt{CARDIoGRAM} are used as the exposure and
outcome datasets. Since the exposure and outcome are the same
variable, it is expected that $\bm{\gamma}$ and $\bm{\Gamma}$ in
our model are the same (or almost equal). Thus $\beta \approx 1$ and $\tau^2
\approx 0$.

In \Cref{tab:cad-cad} we apply our statistical methods to this
validation dataset in three ways: using all the SNPs, using SNPs that
are genome-wide significant in \texttt{MI (UKB)} ($p$-value $\le 5
\times 10^{-8}$), and using SNPs that are not genome-wide significant
in \texttt{MI (UKB)}. In all cases the RAPS point estimates are close to the
truth $\beta = 1$. When only the strong instruments
are used so there is virtually no shrinkage, the shrinkage estimators
are essentially the same as the non-shrinkage estimators. When all the
SNPs are used, the shrinkage estimators are about 10\% more
efficient. MR-Egger and
weighted median are heavily biased by the weak instruments in the
CAD-CAD study. The SIMEX correction \cite{bowden2016assessing}
reduces but does not eliminate the bias of MR-Egger.

\section{Application to the effect of BMI and blood lipids on cardiovascular
  diseases}
\label{sec:appl-caus-effect}

\begin{figure}[tbp]
  \includegraphics[width = \textwidth]{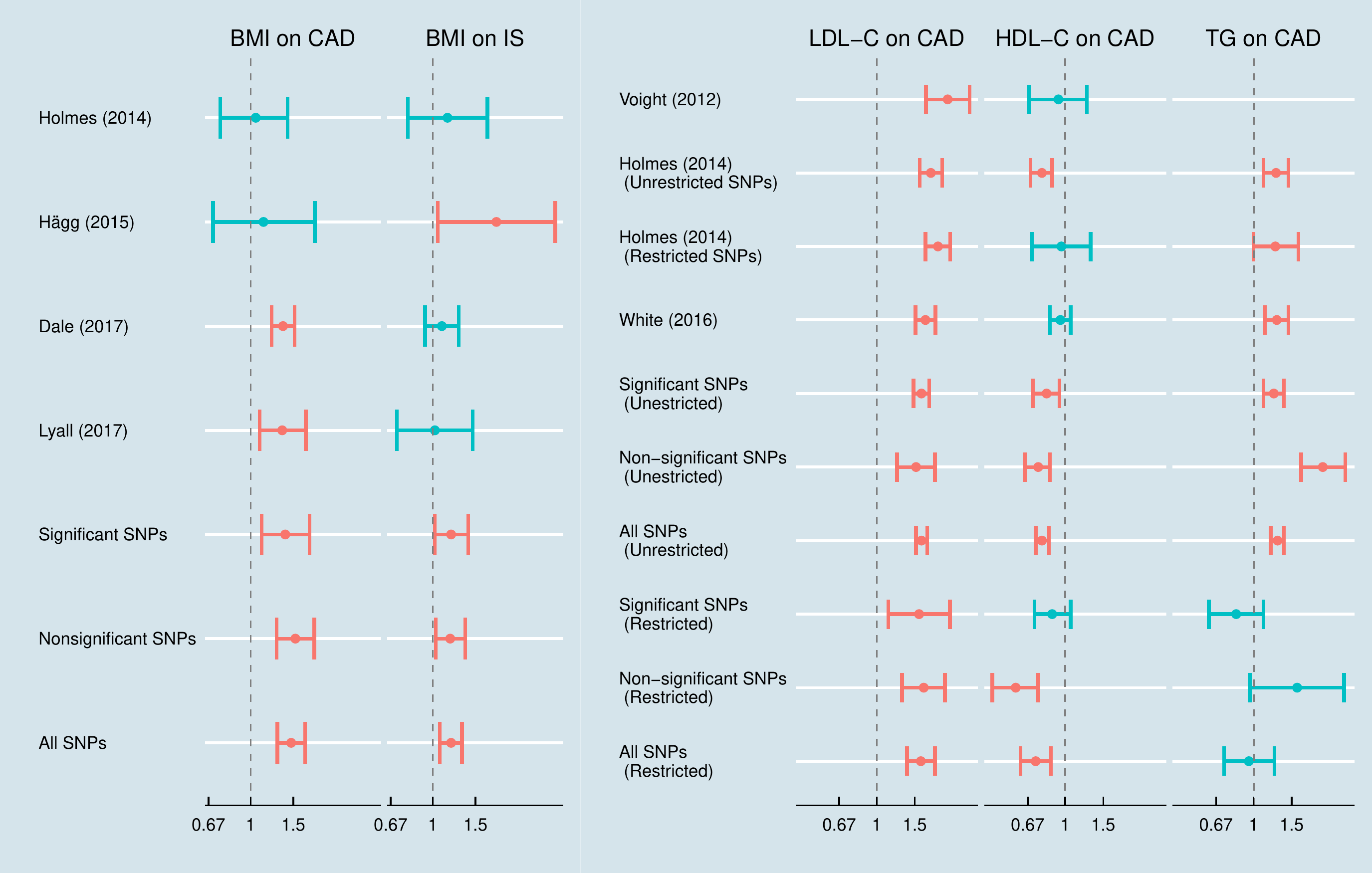}
  \caption{Overview of the MR results for the effect of BMI and blood
    lipids on cardiovascular diseases.}
  \label{fig:ije-results}
\end{figure}

We apply our method to estimate the causal effect of BMI and blood
lipids on cardiovascular disease outcomes. \Cref{fig:ije-results}
summarizes the main results and
\Cref{tab:bmi-results,tab:primary-summary} contain more details. By
default, we use the RAPS estimator with shrinkage weights,
overdispersion adjustment and Huber's score function. In
\Cref{tab:bmi-detailed,tab:lipid1-detailed,tab:lipid2-detailed,tab:lipid3-detailed,tab:lipid4-detailed}
in the Appendix, we report the results using different sets of
instruments and different MR methods (including different specifications
of our RAPS estimator).

\subsection{Body Mass Index (BMI)}
\label{sec:bmi-results}

For the BMI studies we use the \texttt{BMI (Jap)} to select
instruments and the \texttt{BMI (UKB)} to obtain the SNP-exposure
effects. We consider two outcomes, coronary artery disease (CAD) and
ischemic stroke (IS), using two GWAS summary datasets, \texttt{CAD}
and \texttt{IS}, as described in \Cref{tab:list}. The results of the
BMI studies are reported in \Cref{tab:bmi-results}. In summary we find
strong evidence that BMI causally increases the risk of CAD and IS
and the estimated effects remain stable regardless of the instrument
strength. For CAD, in our most powerful analysis using all the
instruments, the estimated odds ratio is $1.47$ (95\% CI 1.29--1.68,
$p$-value $<10^{-7}$). Using the non-significant SNPs reduces the
length of the CI by about 40\%. For IS, the corresponding result is
1.19 (95\% CI 1.07--1.32, $p$-value $=0.0008$), which is much
more precise than the results of previous MR studies.

\begin{table} \centering \setlength{\tabcolsep}{4pt}
  \begin{tabular}[l]{llcc}
    \toprule
    & &  Outcome: CAD  & Outcome: IS \\
    \midrule
    \multicolumn{4}{l}{\textbf{\normalsize Previous MR studies for BMI}}
    \\
    \multicolumn{2}{l}{Holmes et al.\ (2014) \cite{holmes2014causal}$^{*}$ (14 SNPs)} &
    {\color{blue}1.05 (0.75, 1.42)} & {\color{blue}1.15 (0.79, 1.68)} \\
    \multicolumn{2}{l}{H{\"a}gg et al.\ (2015) \cite{hagg2015adiposity} (32 SNPs)} &
    {\color{blue}1.13 (0.70, 1.84)} & 1.83 (1.05, 3.20) \\
    \multicolumn{2}{l}{Dale et al.\ (2017) \cite{dale2017causal} (97 SNPs)} &
    1.36 (1.22, 1.52) & {\color{blue}1.09 (0.93, 1.28)} \\
    \multicolumn{2}{l}{Lyall et al.\ (2017)
      \cite{lyall2017association} (93 SNPs)} & 1.35 (1.09, 1.69) &
    {\color{blue}1.02 (0.71, 1.46)} \\
    \midrule
    \multicolumn{4}{l}{\textbf{\normalsize New MR analysis for BMI}} \\
    \multirow{4}{*}{MR-RAPS} & Using 44 significant SNPs & 1.39 (1.11, 1.75) & 1.19 (1.02, 1.4) \\
    & Using 1075 non-significant SNPs & 1.53 (1.28, 1.83) & 1.18 (1.03, 1.36) \\
    & Using all 1119 SNPs & 1.47 (1.29, 1.68)
    & 1.19 (1.07, 1.32) \\
    & $p$-value for heterogeneity & 0.79 & 0.78 \\
    \cline{1-2}
    \multirow{3}{*}{MR-Egger} & Using 44 significant SNPs & 1.67 (1.16, 2.40) &
    {\color{blue}1.21 (0.95, 1.54)}\\
    & Using 1075 non-significant SNPs & 1.56 (1.27, 1.91) & {\color{blue}1.11
    (0.95, 1.28)} \\
    & Using all 1119 SNPs & 1.47 (1.26,
    1.71) & 1.14 (1.02, 1.27) \\
    \cline{1-2}
    \multirow{3}{*}{Wtd.\ Med.} & Using 44 significant SNPs & 1.32 (1.02, 1.70) & {\color{blue}1.12 (0.91, 1.38)} \\
    & Using 1075 non-significant SNPs & 1.42 (1.15, 1.75) & 1.23 (1.03, 1.46) \\
    & Using all 1119 SNPs & 1.33 (1.09,
    1.62) & {\color{blue}1.12 (0.94, 1.32)} \\
    \bottomrule
  \end{tabular}
  \caption{Previous and new results of the effect of Body Mass Index
    (BMI) on coronary artery disease. {\color{blue} Statistically
    non-significant results are shown in blue color}.\\
    $^{*}$ The original results were reported per 1 kg/m$^2$ of increase in BMI. We transformed the results to per 1 SD increase of BMI
    using SD(BMI) = 4.6 kg/m$^2$.}
  \label{tab:bmi-results}
\end{table}


\subsection{Blood lipids}
\label{sec:results}

For the lipid studies we use the 2010 GWAS reported by Teslovich et
al.\ \cite{teslovich2010biological} to select instruments and the 2013
GWAS (Metabochip data only) by the Global Lipids Genetics Consortium
\cite{willer2013discovery}. We use two independent datasets,
\texttt{CAD} and \texttt{MI (UKB)}, to examine if the MR results
replicate in two studies. Because many SNPs are associated with more
than one lipid traits, we also consider the restricted sets of
instruments that are not associated with the exposure being
studied. For example, a restricted IV for HDL-C must be unassociated
with LDL-C and TG ($p$-value $>0.5$ in the \texttt{screening
  dataset}). Our results for the blood lipids are reported in
\Cref{tab:primary-summary}.

\begin{table} \centering \setlength{\tabcolsep}{4pt} \small
  \begin{tabular}[l]{lccc}
    \toprule
    & Exposure: LDL-c  &  Exposure: HDL-c  & Exposure: TG \\
    \midrule
    \multicolumn{4}{l}{\textbf{\normalsize Observational studies}}  \\
    \multicolumn{4}{l}{Angelantonio et al.\ (2009)
      \cite{assessment2009major}} \\
    \quad\textit{Adjust for non-lipid factors} & {\color{black} 1.56 (1.47, 1.66)}
           & {\color{black} 0.71 (0.68, 0.75)} & {\color{black} 1.37 (1.31, 1.42)} \\
    \quad\textit{Also adjust for lipids} & {\color{black} 1.50 (1.39, 1.61)}
           & {\color{black} 0.78 (0.74, 0.82)} & {\color{blue} 0.99 (0.94, 1.05)} \\
    Voight et al.\ (2012) \cite{voight2012plasma} & {\color{black} 1.54 (1.45, 1.63)} & {\color{black} 0.62
                                                                 (0.58, 0.66)}
                   & {\color{black} 1.42 (1.31, 1.52)} \\
    \midrule
    \multicolumn{4}{l}{\textbf{\normalsize Previous MR studies}} \\
    Voight et al.\ (2012) \cite{voight2012plasma} & {\color{black} 2.13 (1.69, 2.69)} & {\color{blue} 0.93
                                                                 (0.68, 1.26)}
                   & {\color{blue} Not reported}
    \\
    \multicolumn{4}{l}{Holmes et al.\ (2014)
      \cite{holmes2014mendelian}$^{*}$} \\
    \quad \textit{Unrestricted instruments} & {\color{black} 1.78
                                                       (1.58, 2.01)} &
                                                                      {\color{black} 0.78
                                                                      (0.69, 0.87)}
                   & {\color{black} 1.27 (1.11, 1.45)} \\
    \quad \textit{Restricted instruments $(p > 0.01)$} & {\color{black} 1.92
                                                       (1.68, 2.19)} &
                                                                      {\color{blue} 0.96
                                                                      (0.70, 1.31)}
                   & {\color{black} 1.26 (1.00, 1.61)} \\
    White et al.\ (2016) \cite{white2016association} & & & \\
    \quad \textit{MR-Egger} & {\color{black} 1.68 (1.51, 1.87)} & {\color{blue} 0.95
                                                        (0.85, 1.06)} &
    {\color{black} 1.28 (1.13, 1.45)} \\
    \quad \textit{Multivariable MR} & {\color{black} 1.50 (1.39, 1.63)} & {0.86
                                                        (0.78, 0.96)} &
    {\color{black} 1.38 (1.19, 1.59)} \\
    \midrule
    \multicolumn{4}{l}{\textbf{\normalsize New MR analysis: Lipids-CAD
      (CARDIoGRAMplusC4D)}}
    \\
    \multicolumn{4}{l}{\textit{Unrestricted instruments}} \\
    Using significant SNPs & {\color{black} 1.61 (1.48, 1.75)} & {\color{black} 0.82 (0.71, 0.94)} &
    {\color{black} 1.24 (1.11, 1.38)}\\
    Using non-significant SNPs & {\color{black} 1.52 (1.24, 1.86)} & {\color{black} 0.75 (0.65, 0.85)}
                       & {\color{black} 2.09 (1.66, 2.66)} \\
    Using all SNPs & {\color{black} 1.61 (1.52, 1.71)} & {\color{black} 0.78 (0.73, 0.84)} & {\color{black} 1.29
                                                       (1.2, 1.38)} \\
    $p$-value for heterogeneity & 0.21 & $<0.001$ & $<0.001$ \\
    \cline{1-1}
    \multicolumn{4}{l}{\textit{Restricted instruments ($p > 0.5$)}} \\
    Using significant SNPs & {\color{black} 1.57 (1.13, 2.18)} & {\color{blue} 0.87 (0.72, 1.06)} &
    {\color{blue} 0.83 (0.62, 1.11)}\\
    Using non-significant SNPs & {\color{black} 1.65 (1.31, 2.07)} & {\color{black} 0.59 (0.46, 0.75)}
                       & {\color{blue} 1.59 (0.96, 2.62)} \\
    Using all SNPs & {\color{black} 1.6 (1.38, 1.86)} & {\color{black} 0.73 (0.62, 0.86)} & {\color{blue} 0.95
                                                       (0.73, 1.25)}
                                                     \\
    $p$-value for heterogeneity & 0.08 & 0.36 & 0.19 \\
    \midrule
    \multicolumn{4}{l}{\textbf{\normalsize New MR analysis: Lipids-MI (UK BioBank)}}
    \\
    \multicolumn{4}{l}{\textit{Unrestricted instruments}} \\
    Using significant SNPs & {\color{black} 1.27\% (1.04\%, 1.49\%)} & {\color{blue} -0.43\% (-0.96\%, 0.10\%)} &
    {\color{black} 0.72\% (0.32\%, 1.12\%)}\\
    Using non-significant SNPs & {\color{black} 1.06\% (0.39\%, 1.72\%)} & {\color{black} -1.07\% (-1.54\%, -0.60\%)}
                       & {\color{black} 2.41\% (1.61\%, 3.21\%)} \\
    Using all SNPs & {\color{black} 1.24\% (1.03\%, 1.45\%)} & {\color{black} -0.77\% (-1.02\%, -0.53\%)} & {\color{black} 1.03\%
                                                       (0.77\%,
                                                       1.28\%)} \\
    $p$-value for heterogeneity & 0.99 & $<0.001$ & $<0.001$ \\
    \cline{1-1}
    \multicolumn{4}{l}{\textit{Restricted instruments ($p > 0.5$)}} \\
    Using significant SNPs & {\color{black} 1.37\% (0.25\%, 2.49\%)} & {\color{blue} -0.25\% (-1.04\%, 0.53\%)} &
    {\color{blue} 0.36\% (-0.77\%, 1.49\%)}\\
    Using non-significant SNPs & {\color{black} 1.49\% (0.53\%, 2.45\%)} & {\color{black} -2.26\% (-3.14\%, -1.37\%)}
                       & {\color{blue} 1.7\% (-0.61\%, 4.00\%)} \\
    Using all SNPs & {\color{black} 1.25\% (0.63\%, 1.88\%)} & {\color{black} -1.28\% (-1.88\%, -0.67\%)} & {\color{blue} 0.51\%
                                                       (-0.55\%,
                                                       1.57\%)} \\
    $p$-value for heterogeneity & 0.66 & 0.1 & 0.06 \\
    \bottomrule
  \end{tabular}
  \caption{\small Previous and new results of the effect of major blood
    lipids on coronary artery disease (or myocardial infarction) risk.
    The numbers reported in the Angelantonio et al.\ (2009) study are
    hazard ratios, and the numbers reported in the new MR analysis
    using UK BioBank are risk differences. All other numbers are odds
    ratios. The prevalence of myocardial infarction in UK BioBank is
    $8288/360420 = 2.30\%$, so a risk difference of $1\%$ roughly
    corresponds to an odds ratio of $1.45$. {\color{blue} Statistically
    non-significant results are shown in blue color}.  \\
  $^{*}$ The original results for the Holmes et al.\ (2014 study) were reported per
  1 mmol/L increase of LDL-c and HDL-c and 1 log unit increase of
  TG. We transformed the results to per 1 SD increase using the
  following approximate standard deviations: SD(LDL-c) = 1
  mmol/L, SD(HDL-c) = 0.4 mmol/L, SD(Log TG) = 0.5. \\
}
  \label{tab:primary-summary}
\end{table}


Similar to BMI, the results for LDL-C are highly significant and
stable across different datasets and sets of instruments. For CAD, in our
most accurate result using all the unrestricted instruments, the
estimated odds ratio is 1.61 (95\% CI 1.52--1.71). The CI is much shorter than
previous MR studies. The estimated odds ratio does not move by much
when we use the restricted instruments: 1.6 (95\% CI 1.38--1.86),
though the CI becomes wider because fewer SNPs are used. Similar
observations are found when \texttt{MI (UKB)} is used as the outcome.

For HDL-C, the MR analyses using all the SNPs appears to suggest that
HDL-C is protective. For example, in the analysis using the
\texttt{CAD} dataset and all the unrestricted instruments, the
estimated odds ratio is 0.78 (95\% CI 0.73--0.84). However, this
highly significant result is mostly driven by the weaker
instruments. When using the SNPs that are not genome-wide significant
in \texttt{HDL (2010)} dataset, the estimated causal effect becomes
weaker and not statistically significant when using the restricted
instruments or the \texttt{MI (UKB)} data. The diagnostic plots
(\Cref{fig:hdc-diagnostics-1,fig:hdc-diagnostics-2,fig:hdc-diagnostics-3,fig:hdc-diagnostics-4})
also show strong evidence of heterogeneity between the strong and weak
instruments.

For TG, the MR analyses using unrestricted instruments all yield
statistically significant results. When using the \texttt{CAD} dataset
and all the unrestricted instruments, the estimated odds ratio is 1.29
(95\% CI 1.2--1.38). However, our heterogeneity test shows strong evidence of
heterogeneity. When using the restricted instruments for both studies,
the confidence intervals become very wide and not statistically
significant due to the lack of instruments that are exclusively
associated with TG. In conclusion, the potential causal role of
triglycerides in CAD remains uncertain based on our results.

\section{Discussion}
\label{sec:discussion}

Our examples in \Cref{sec:validation-studies,sec:appl-caus-effect}
demonstrate that a genome-wide MR design is usually much more powerful than a
MR analysis that just uses a small set of strong genetic
instruments. The empirical partially Bayes technique introduced in
this paper can further increases the statistical power.

Our empirical results reaffirm the causal effect of adiposity and
LDL-C on the risk of coronary artery disease. Additionally, our MR
analysis gives strong support for the causal role of adiposity in the
development of ischemic stroke. Another observation is that in these
cases the estimated causal effects are very close across different
strength of the instruments. The homogeneity of the effect further
adds evidence to these causal relationships.

In comparison, although our most powerful genome-wide MR analyses
show that the effect of HDL-C on CAD is highly significant,
there is also strong evidence of effect heterogeneity. Indeed, the statistical
significance is mostly driven by the weak instruments. When used
alone, the weak instruments give very different effect estimates (with
non-overlapping confidence intervals)  than the strong instruments. The
role of HDL-C in cardiovascular disease has been heatedly debated in
the recent years following the failure of several highly anticipated
clinical trials for the \emph{CETP} inhibitors
\cite{barter2007effects,schwartz2012effects,lincoff2017evacetrapib}. Observational
epidemiology studies have long suggested that HDL-C is
inversely associated with the risk of myocardial infarction
\cite{assessment2009major,miller1975plasma,prospective2007blood}. However,
the failed \emph{CETP} trials and previous MR studies
\cite{voight2012plasma,holmes2014causal} have led to the broad
conclusion that HDL-C is unlikely a causal agent for atherosclerotic
cardiovascular disease \citep{holmes2017mendelian,rosenson2018hdl},
though some remain hopeful in the HDL function hypothesis
\citep{rosenson2018hdl,rader2014hdl,rohatgi2014hdl}. Our MR analyses
suggest that the causal role of HDL-C remains uncertain and the
effect of HDL-C is heterogeneous using different
instruments. Relatedly, Bulik-Sullivan et al.\ (2015)
\cite{bulik2015atlas} also observed statistical significant genetic
correlation (computed across the whole genome) between HDL-C and CAD
(see also \Cref{fig:ld-score}). We think further investigations are
needed to demystify the strong observational and genetic associations
between HDL-C and CAD.


\bibliographystyle{vancouver}
\bibliography{ref}

\newpage
\appendix
\renewcommand{\thetable}{\Alph{section}\arabic{table}}
\renewcommand{\thefigure}{\Alph{section}\arabic{figure}}

\section{Technical details}
\label{sec:technical-details}

\setcounter{table}{0}
\setcounter{figure}{0}

\subsection{The empirical partially Bayes approach}
\label{sec:empir-part-bayes}

We propose a new way of eliminating the nuisance
parameters $\bm \gamma$ that can increase the power of
genome-wide MR studies when most IVs are
very weak (i.e.\ $\bm \gamma$ are very close to $0$). The key idea is to view the
errors-in-variables regression as a semiparametric problem: instead of
treating the vector $\bm \gamma$ as the nuisance parameter whose
dimension grows as more SNPs are used, we treat the (empirical)
distribution of $\bm \gamma$ as the nuisance. This idea originates
from the general solution given by Kiefer and Wolfowitz \cite{kiefer1956consistency} to
the Neyman-Scott problem in which the observed data
are modeled by a mixture distribution. In principle, the
statistical inference can be carried out by solving a nonparametric
maximum likelihood problem, but the numerical problem is often
extremely challenging \cite{feng2018approximate}.

We will take an empirical partially Bayes
approach introduced by Lindsay \cite{lindsay1985using} which is numerically feasible and
still has several good theoretical properties. The approach is
partially Bayes because only the nuisance parameters are modeled by a
prior distribution \cite{cox1975note,meng2010automated}. It is
empirical Bayes because the prior distribution is estimated
empirically using the observed data.

Consider the
simplest scenario where $\bm \alpha = \bm 0$ and derive the
conditional score function \cite{lindsay1982conditional}. When $\bm
\alpha = \bm 0$, the log-likelihood function of the data $(\hat{\bm
  \gamma},\hat{\bm \Gamma})$ is given by
\[
  l(\beta, \bm \gamma) = \prod_{j=1}^p
  l_j(\beta,\gamma_j),~\text{where}~l_j(\beta,\gamma_j)= - \frac{(\hat
    \gamma_j - \gamma_j)^2}{2 \sigma_{Xj}^2} - \frac{(\hat \Gamma_j - \beta\gamma_j)^2}{2 \sigma_{Yj}^2}.
\]
Thus the score function of $\beta$ in the $j$-th SNP is given by
\[
  S_j(\beta,\gamma_j) = \frac{\partial}{\partial \beta}
  l_j(\beta,\gamma_j) =
  \frac{\gamma_j(\hat{\Gamma}_j - \beta \gamma_j)}{\sigma_{Yj}^2}.
\]
and a sufficient statistic of the nuisance parameter $\gamma_j$ is
\begin{equation} \label{eq:gamma-suff-stat}
  W_j(\beta) = \frac{\hat{\gamma}_j}{\sigma_{Xj}^2} + \frac{\beta \hat \Gamma_j}{\sigma_{Yj}^2}.
\end{equation}
When $\beta$ is given, the maximum likelihood estimator (MLE) of
$\gamma_j$ is
\begin{equation*} \label{eq:mle}
  \hat{\gamma}_{j,\mathrm{MLE}}(\beta) = \frac{W_j(\beta)}{1 / \sigma_{Xj}^2
    + \beta^2 / \sigma_{Yj}^2}.
\end{equation*}

The conditional score function is the residual of the score function $S_j$
conditioning on $W_j$. After some algebra, we obtain
\begin{equation} \label{eq:conditional-score}
  C_j(\beta,\gamma_j) = S_j(\beta,\gamma_j) - \mathrm{E}\left[
    S_j(\beta,\gamma_j) | W_j(\beta) \right] =
  \frac{\gamma_j(\hat{\Gamma}_j - \beta \hat{\gamma}_j)}{\beta^2
    \sigma_{Xj}^2 + \sigma_{Yj}^2}.
\end{equation}
We would like to make three remarks on the conditional score
\eqref{eq:conditional-score}. First, it is proportional to the
``regression residual'' $\hat{\Gamma}_j - \beta \hat{\gamma}_j$ which
has mean 0 at the true $\beta$. Second, the nuisance parameter appears
in \eqref{eq:conditional-score} only as a weight to the ``regression
residual'', as noticed by Lindsay \cite{lindsay1985using}. Third, the sufficient statistic
$W_j(\beta)$ of $\gamma_j$ in \eqref{eq:gamma-suff-stat} is
independent of $\hat{\Gamma}_j - \beta \hat{\gamma}_j$ because they
are jointly normal and their covariance is $0$, regardless of what
$\beta$ is. See \cite{stefanski1987conditional} for a related
application of the conditional score in measurement error models.
These observations motivate the following estimating
function of $\beta$:
\begin{equation} \label{eq:estimating-function}
  C(\beta) = \sum_{j=1}^p
  C_j\Big(\beta,\hat{\gamma}_j\big(\beta,W_j(\beta)\big)\Big) =
  \sum_{j=1}^p \frac{\hat{\gamma}_j\big(\beta, W_j(\beta)\big) \cdot
    (\hat{\Gamma}_j - \beta \hat{\gamma}_j)}{\beta^2 \sigma_{Xj}^2 +
    \sigma_{Yj}^2},
\end{equation}
where $\hat{\gamma}_j\big(\beta, W_j(\beta)\big)$ is any estimator of
$\hat{\gamma}_j$ that only depends on the sufficient statistic
$W_j(\beta)$, not necessarily the MLE. It is obvious that the
estimating function is always unbiased,
i.e.\ $\mathbb{E}[C(\beta)] = 0$ at the true value of $\beta$,
regardless of what form of $\hat{\gamma}_j$ is used. The estimator
$\hat{\beta}$ is obtained by solving $C(\beta) = 0$.

The profile score approach \cite{zhao2018statistical} can be
viewed as a special case of the conditional score, where the
``weights'' are the MLE of $\gamma_j$: $\hat{\gamma}_j\big(\beta,
W_j(\beta)\big) = \hat{\gamma}_{j,\mathrm{MLE}}(\beta)$. This is also
equivalent to using a flat prior in the partially Bayes approach that
will be explained shortly. Under
regularity conditions, we prove in our previous article \cite{zhao2018statistical} that the
profile score provides a consistent and asymptotically normal
estimator of $\beta$. However, this estimator is
not efficient in general. To see this, let's assume most $\gamma_j$ are equal to
$0$. Intuitively, the $j$-th IV provides no information on
$\beta$ because the distribution of $(\hat{\gamma}_j,\hat{\Gamma}_j)$
does not depend on $\beta$. However, some information is still used to
estimate $\gamma_j$ in the MLE, resulting in a loss of statistical
efficiency. This phenomenon is particularly relevant in genome-wide MR
as most of the genetic instruments are very weak.

When $\sigma_{Xj}^2$ and $\sigma_{Yj}^2$ are equal across $j$,
Lindsay \cite{lindsay1985using} points out that the efficient estimator of $\beta$ is
given by the weight $\hat{\gamma}_j^{*} =
\mathbb{E}_{\pi^{*}}[\gamma_j|W_j(\beta)]$, where $\pi^{*}$ is the
empirical distribution of $\bm{\gamma}$. However, since $\bm{\gamma}$
and hence $\pi^{*}$ is unknown, it is impossible to compute
$\hat{\gamma}_j^{*}$ directly. Lindsay proposes to
use the empirical Bayes (EB)
estimator of $\bm \gamma$. Suppose the distribution of $\bm \gamma$ is
modeled by a parametric family $\pi_{\eta}$ and $\hat{\eta}$ is an
estimate of $\eta$ using the observed data. We can use
\begin{equation} \label{eq:gamma-eb}
  \hat{\gamma}_{j,\mathrm{EB}}(\beta, W_j(\beta)) =
  \mathbb{E}_{\pi_{\hat{\eta}}}[\gamma_j|W_j(\beta)]
\end{equation}
in the estimating function \eqref{eq:estimating-function}. Since this
is usually a better estimator of the whole vector $\bm \gamma$ than
the MLE, a phenomenon known as the James-Stein paradox
\cite{james1961estimation,efron1973stein}, it is natural to expect
that the resulting function of $\beta$ is also more efficient than the
profile score. In
fact, Lindsay \cite{lindsay1985using} shows that the estimator has a local
efficiency property: when the parametric distribution $\pi_{\eta}$ is
specified correctly, the estimator $\hat{\beta}$ is asymptotically
efficient; when $\pi_{\eta}$ is specified incorrectly, the estimator
is not efficient but still consistent.


\subsection{Implementation details}
\label{sec:impl-deta}

\subsubsection{Spike-and-slab prior}

Model \eqref{eq:spike-slab} implies that
$\hat{\gamma}_j/\sigma_{Xj}$ also follows a Gaussian mixture
distribution marginally:
\begin{equation} \label{eq:marginal-model}
  \hat{\gamma}_j/\sigma_{Xj} \overset{i.i.d.}{\sim} p_1 \cdot \mathrm{N}(0, \sigma_1^2+1)
  + (1 - p_1) \cdot \mathrm{N}(0, \sigma_2^2 + 1),~j=1,\dotsc,p.
\end{equation}
In practice we use maximum likelihood to estimate the prior parameters
$(p_1,\sigma_1^2,\sigma_2^2)$ by fitting the marginal
mixture model \eqref{eq:marginal-model} to the exposure $z$-statistics
$\hat{\gamma}_j/\sigma_{Xj}$, $j=1,\dotsc,p$.

The posterior mean of $\gamma_i/\sigma_{Xi}$ can be computed using the
formulas in \Cref{prop:posterior}.
\begin{proposition} \label{prop:posterior}
  Suppose $Z \sim \mathrm{N}(\gamma,\sigma^2)$, $\gamma \sim p_1
  \mathrm{N}(\mu_1,\sigma_1^2) + (1-p_1) \mathrm{N}(\mu_2,\sigma_2^2)$,
  then $\gamma|Z \sim \tilde{p} \cdot
  \mathrm{N}(\tilde{\mu}_1,\tilde{\sigma}_1^2) + (1 - \tilde{p}) \cdot
  \mathrm{N}(\tilde{\mu}_2,\tilde{\sigma}_2^2)$, where
  \[
    \tilde{\mu}_k = \frac{Z/\sigma^2 + \mu_k/\sigma_k^2}{1/\sigma^2 +
      1/\sigma_k^2},~\tilde{\sigma}_k^2 = \frac{1}{1/\sigma^2 +
      1/\sigma_k^2},~\text{and}
  \]
  \[
    \tilde{p} = \frac{p_1 \cdot \varphi(Z;\mu_1,\sigma^2+\sigma_1^2)}{p_1 \cdot
      \varphi(Z;\mu_1,\sigma^2+\sigma_1^2) + (1-p_1)\cdot
      \varphi(Z;\mu_2,\sigma^2+\sigma_2^2)}.
  \]
  In the above equation, $\varphi(z;\mu,\sigma^2)$ is the probability
  density function of the normal distribution $\mathrm{N}(\mu,\sigma^2)$:
  $\varphi(z;\mu,\sigma^2) = (\sqrt{2 \pi \sigma^2})^{-1} \exp\{-(z -
  \mu)^2/(2 \sigma^2)\}$. The posterior mean of $\gamma$ is given by
  $\hat{\gamma} = \mathbb{E}[\gamma|Z] = \tilde{p} \tilde{\mu}_1 + (1 -
  \tilde{p}) \tilde{\mu}_2$.
\end{proposition}

We want to make two remarks about the choice of prior
distribution. First, there is an attractive property of setting the
means to be $0$ in \eqref{eq:spike-slab}. Using \Cref{prop:posterior},
it is easy to verify that, when $\mu_1 = \mu_2 = 0$,
$\mathbb{E}[\gamma|Z] = - \mathbb{E}[\gamma|-Z]$. As a consequence,
the estimating functions in \eqref{eq:eb-estimating-function} are
invariant to allele-recoding, meaning if a pair of observations
$(\hat{\gamma}_j,\hat{\Gamma}_j)$ is replaced by $(- \hat{\gamma}_j,-
\hat{\Gamma}_j)$, the point estimate $\hat{\beta}$ is unchanged. This
is desirable because the allele coding used in a GWAS is often
arbitrary. The second remark is that the spike-and-slab implementation
is actually quite important in order to gain efficiency. To see this,
suppose a single Gaussian prior is used (as in the empirical Bayes
interpretation of the James-Stein estimator). It is easy to show that
every SNP then receives the same amount of multiplicative shrinkage,
so the first estimating function in \eqref{eq:eb-estimating-function}
is just scaled by a constant. As a consequence, the estimator
$\hat{\beta}$ is the same no matter how large the shrinkage is. By
using a spike-and-slab prior, every genetic instrument is shrunken selectively
\cite{ishwaran2005spike} according to the its strength and
thus efficiency might be gained. It is then natural to expect that the
efficiency gain is more substantial when the two components are more
distant ($\sigma_1$ and $\sigma_2$ are more different). See
\Cref{sec:validation-studies} for an example.

\begin{figure}[ht] \renewcommand\thefigure{A1}
  \centering
  \includegraphics[width = \textwidth]{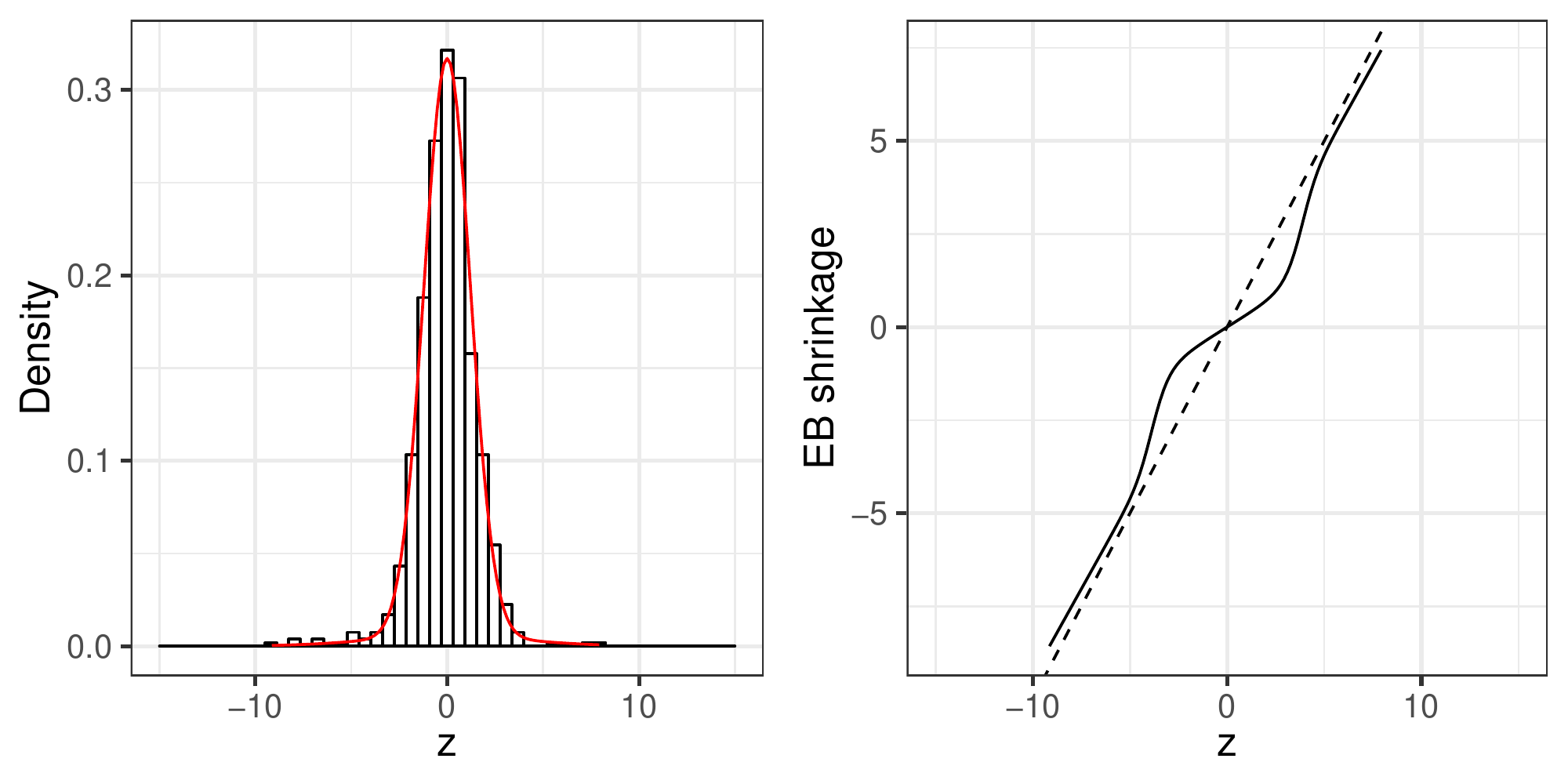}
  \caption{Examine the fitted prior distribution for HDL-C using
    $861$ restricted SNPs used in \Cref{tab:lipid3-hdl}. The left panel
    compares the empirical distribution of $z_j =
    \hat{\gamma}_j/\sigma_{Xj}$ (black histogram) with the fitted
    Gaussian mixture distribution in \eqref{eq:marginal-model} (red
    curve). The right panel shows the posterior mean as a function of
    the $z$-score.}
  \label{fig:prior-hdl}
\end{figure}

In \Cref{fig:prior-hdl} we examine the fit of the Gaussian mixture
model \eqref{eq:marginal-model} for our primary analysis of HDL-c in
\Cref{sec:appl-caus-effect}. In this example we selected $1122$
SNPs and the estimated prior parameters are $p_1 = 0.91$, $\sigma_1 =
0.73$, $\sigma_2 = 4.57$. In the left panel of \Cref{fig:prior-hdl} we
compare the empirical distribution of $\hat{\gamma}_j/\sigma_{Xj}$
(black histogram) with the fitted Gaussian mixture distribution in
\eqref{eq:marginal-model} (red curve). We find the empirical fit is
quite good. In the right panel of \Cref{fig:prior-hdl} we plot the
empirical Bayes shrinkage estimator as a function of the
$z$-score. When the $z$-score is close to $0$, it is shrunken
aggressively towards $0$; when the $z$-score is large (e.g.\ greater
than $5$), there is essentially no shrinkage.

\subsubsection{Choice of robust score function}
\label{sec:choice-robust-score}

In our empirical analysis we will consider two choices of the
function $\psi(\cdot)$. The first is the identity function $\psi_I(t) =
t$ which is non-robust, and the second is the Huber score function
\cite{huber1964robust} that is robust:
\[
  \psi_H(t) =
  \left\{
    \begin{array}{ll}
      t, & \text{if}~|t| \le k, \\
      k \cdot \mathrm{sign}(t), & \text{otherwise}.
    \end{array}\right.
\]
The tuning constant $k$ is chosen to be $1.345$, which corresponds to $95\%$
asymptotic efficiency for normal samples in the standard location
problem. Another common choice of the robust score function is Tukey's
biweight \cite{maronna2006robust}, where large outliers essentially
have no influence on the estimator. In practice we find that using
Tukey's biweight usually gives more local roots than Huber's
score. Thus we only report results of the more stable Huber's score
in the application.

\subsubsection{Multiple roots of the estimating equations}
\label{sec:mult-roots-estim}

In practice, the estimating functions in \eqref{eq:eb-estimating-function}
may have multiple roots. Some roots are trivial: it is straightforward to show
that $\tilde{\bm C}(\beta,\tau^2) \to 0$ if $\beta \to \pm \infty$ or
$\tau^2 \pm \infty$. These unbounded roots can be easily ruled
out. However, often there are multiple finite roots. When this
happens, we report the root whose $\hat{\beta}$ is closest to the
profile-likelihood estimator of $\beta$ assuming all the SNPs are
valid IVs \cite{zhao2018statistical}. The latter is always unique
because it solves an optimization problem. When there is another root
that is also
close to the profile-likelihood estimator (the criterion we use in
the application is the absolute difference is no more than 5 times the
closest difference), we report the empirical partially Bayes estimator
is not available.

\subsubsection{Standard error of the estimator}
\label{sec:variance-estimator}

Our final problem is to compute the standard error of the estimator
$\hat{\bm \theta} = (\hat{\beta},\hat{\tau}^2)$. For the calculation
below we assume all SNPs satisfy $\alpha_j \sim \mathrm{N}(0,
\tau^2)$, i.e.\ there is no outlier. After taking a first-order Taylor
expansion at the true value of $\bm \theta = (\beta,\tau^2)$,
\[
  \bm 0 = \tilde{\bm C}(\hat{\bm \theta}) \approx \tilde{\bm
    C}(\bm \theta) + \nabla \tilde{\bm C} (\bm \theta) \cdot (\hat{\bm
    \theta} - \bm \theta),
\]
the variance of $\hat{\bm \theta}$ can be approximated using the Delta
method by
\[
  \mathrm{Var}(\hat{\bm \theta}) \approx \Big[\nabla \tilde{\bm C} (\bm
  \theta)\Big]^{-1} \mathrm{Var}\left(\tilde{\bm C}(\bm
    \theta)\right) \Big[\nabla \tilde{\bm C} (\bm
  \theta)\Big]^{-T}.
\]
By repeatedly using the fact that
$\hat{\gamma}_{j,\mathrm{EB}}$ is independent of $t_j(\beta,\tau^2)$
(so several terms have mean $0$ and are dismissed), we obtain, after some
algebra, that
\begin{equation} \label{eq:observed-information}
  \begin{split}
    \mathrm{Var}\left(\tilde{\bm C}(\bm
      \theta)\right) &\approx
    \sum_{j=1}^p \begin{pmatrix}
      c_1
      \hat{\gamma}_j^2 \big/ s_j^2 & 0 \\
      0 & c_2
      \hat{\gamma}_j^2 \big/ s_j^4
    \end{pmatrix},~\text{and} \\
    \nabla \tilde{\bm C}(\bm \theta) &\approx
    \sum_{j=1}^p
    \begin{pmatrix}
      \left[\psi_1(t_j) \cdot (\partial/\partial \beta) \hat{\gamma}_j +
        \hat{\gamma}_j \psi_1'(t_j) \cdot (\partial/\partial \beta) t_j
      \right] / s_j &   \left[\psi_1(t_j) \cdot (\partial/\partial \tau^2) \hat{\gamma}_j \right] / s_j \\
      0 & (\delta + c_3) / (2 s_j^4) \\
    \end{pmatrix}.
  \end{split}
\end{equation}
The subscript EB and the dependence of $s_j$ and $t_j$ on $\bm \theta$
are suppressed to simplify the expressions. The constants appeared in
\eqref{eq:observed-information} are $\delta = \mathbb{E}[\psi_2(Z)]$,
$c_1 = \mathbb{E}[\psi_1^2(Z)]$, $c_2 = \mathrm{Var}(\psi_2(Z))$, $c_3
= \mathbb{E}[Z \psi'(Z) - \psi(Z)]$ for $Z \sim \mathrm{N}(0,
1)$. Assuming $\hat{\bm \theta}$ is a good estimator of $\bm \theta$,
the matrices in \eqref{eq:observed-information} can be estimated by
replacing $\bm \theta$ with $\hat{\bm \theta}$.

\subsubsection{Diagnostics}
\label{sec:diagnostics}

To check the modeling assumptions we propose to use a scatterplot of
standardized residuals $t_j(\hat{\beta},\hat{\tau}^2)$
versus the empirical Bayes estimates
$\hat{\gamma}_{j,\mathrm{EB}}(\hat{\beta},\hat{\tau}^2)$,
$j=1,\dotsc,p$. Note that both $t_j$ and
$\hat{\gamma}_{j,\mathrm{EB}}$ depend on the allele coding. To ease
the visualization, we choose the allele coding such that
$\hat{\gamma}_{j,\mathrm{EB}}$ is positive.
Under our modeling assumptions, if
$(\hat{\beta},\hat{\tau}^2)$ is close to the true value, most of
$t_j(\hat{\beta},\hat{\tau}^2)$ should be independent of
$\hat{\gamma}_{j,\mathrm{EB}}(\hat{\beta},\hat{\tau}^2)$ and
distributed like a standard normal. We can verify this implication by
computing a smoothing spline of the scatter-plot and check it is
different from the $x$-axis (constant $0$). More specifically, we run
a linear regression with B-splines of $t_j(\hat{\beta},\hat{\tau}^2)$
with degrees of freedom $\lfloor p/50 \rfloor$ and report the $F$-test result
as ``heterogeneity $p$-value''. We also use the Q-Q plot
of $t_j(\hat{\beta},\hat{\tau}^2)$ against standard normal to check if
there is excessive pleiotropy that could not be explained by the normal
random effects model.


\newpage
\section{Data availability}
\label{sec:data-avail}

\setcounter{table}{0}
\setcounter{figure}{0}

\begin{description}
\item[BMI (Jap):] GWAS summary dataset is downloaded from
\url{ftp://ftp.ebi.ac.uk/pub/databases/gwas/summary_statistics/AkiyamaM_28892062_GCST004904}.
\item[BMI (UKB) and MI (UKB):] Round 2 GWAS summary results for the UK
  BioBank data are available at
  \url{http://www.nealelab.is/uk-biobank/}.
\item[LDL-C (2010), HDL-C (2010), and TG (2010):] GWAS summary dataset
  is downloaded from
  \url{http://csg.sph.umich.edu/willer/public/lipids2010/}.
\item[LDL-C (2010), HDL-C (2010), and TG (2010):] GWAS summary results
  for the Metabochip data are downloaded from
  \url{http://csg.sph.umich.edu/willer/public/lipids2013/}.
\item[CAD:] GWAS summary dataset is downloaded from the
  CARDIoGRAMplusC4D Consortium website:
  \url{http://www.cardiogramplusc4d.org/data-downloads/}.
\item[IS:] GWAS summary results for the European samples are
  downloaded from
  \url{ftp://ftp.ebi.ac.uk/pub/databases/gwas/summary_statistics/MalikR_29531354_GCST005843}.
\end{description}

\newpage
\section{Detailed results}
\label{sec:detailed-results}

\setcounter{table}{0}
\setcounter{figure}{0}

\subsection{Results}

In
\Cref{tab:bmi-detailed,tab:lipid1-detailed,tab:lipid2-detailed,tab:lipid3-detailed,tab:lipid4-detailed},
we report detailed results of our MR analyses.

\begin{table}[h] \centering \small
\begin{subtable}[t]{\textwidth} \centering
\begin{tabular}{l|cccccc}
\hline
 & \multicolumn{2}{c}{$p_{\mathrm{sel}} \in (0, 1)$} & \multicolumn{2}{c}{$p_{\mathrm{sel}} \in (0, 5 \times 10^{-8})$} & \multicolumn{2}{c}{$p_{\mathrm{sel}} \in (5 \times 10^{-8}, 1)$} \\

\hline
\# SNPs  &  \multicolumn{2}{c}{1119} &  \multicolumn{2}{c}{44} &  \multicolumn{2}{c}{1075} \\
$p_1$  &  \multicolumn{2}{c}{0.84} &  \multicolumn{2}{c}{0.81} &  \multicolumn{2}{c}{0.52} \\
$\sigma_1$  &  \multicolumn{2}{c}{1.41} &  \multicolumn{2}{c}{6.55} &  \multicolumn{2}{c}{0.69} \\
$\sigma_2$  &  \multicolumn{2}{c}{5.57} &  \multicolumn{2}{c}{14.22} &  \multicolumn{2}{c}{2.82} \\
MR-Egger  &  \multicolumn{2}{c}{0.386 (0.077)} &  \multicolumn{2}{c}{0.513 (0.184)} &  \multicolumn{2}{c}{0.442 (0.105)} \\
Wtd.\ Med.  &  \multicolumn{2}{c}{0.284 (0.1)} &  \multicolumn{2}{c}{0.278 (0.124)} &  \multicolumn{2}{c}{0.348 (0.105)} \\
\hline  & MLE & Shrinkage & MLE & Shrinkage & MLE & \multicolumn{1}{c}{Shrinkage} \\
$\tau_2 = 0$,$\psi_I$  & 0.382 (0.061) & 0.388 (0.06) & 0.291 (0.086) & 0.292 (0.086) & 0.447 (0.084) & 0.454 (0.082) \\
$\tau_2 = 0$,$\psi_H$  & 0.398 (0.061) & 0.401 (0.061) & 0.345 (0.088) & 0.346 (0.088) & 0.446 (0.084) & 0.452 (0.083) \\
$\tau_2 \ne 0$,$\psi_I$  & 0.367 (0.067) & 0.374 (0.066) & 0.297 (0.12) & 0.298 (0.12) & 0.418 (0.09) & 0.427 (0.089) \\
$\tau_2 \ne 0$,$\psi_H$  & 0.382 (0.068) & 0.387 (0.068) & 0.332 (0.117) & 0.332 (0.117) & 0.42 (0.092) & 0.426 (0.091) \\
\hline
\end{tabular}
\caption{Screening: BMI (Dataset: Akiyama et al. (2017)); Exposure:
  BMI (Dataset: UK BioBank); Outcome: CAD, (Dataset: CARDIoGRAMplusC4D).}
\end{subtable}
\begin{subtable}[t]{\textwidth} \centering
\begin{tabular}{l|cccccc}
\hline
 & \multicolumn{2}{c}{$p_{\mathrm{sel}} \in (0, 1)$} & \multicolumn{2}{c}{$p_{\mathrm{sel}} \in (0, 5 \times 10^{-8})$} & \multicolumn{2}{c}{$p_{\mathrm{sel}} \in (5 \times 10^{-8}, 1)$} \\

\hline
\# SNPs  &  \multicolumn{2}{c}{1880} &  \multicolumn{2}{c}{63} &  \multicolumn{2}{c}{1817} \\
$p_1$  &  \multicolumn{2}{c}{0.88} &  \multicolumn{2}{c}{0.81} &  \multicolumn{2}{c}{0.77} \\
$\sigma_1$  &  \multicolumn{2}{c}{1.2} &  \multicolumn{2}{c}{5.38} &  \multicolumn{2}{c}{0.94} \\
$\sigma_2$  &  \multicolumn{2}{c}{5.35} &  \multicolumn{2}{c}{13.05} &  \multicolumn{2}{c}{3.07} \\
MR-Egger  &  \multicolumn{2}{c}{0.13 (0.056)} &  \multicolumn{2}{c}{0.193 (0.122)} &  \multicolumn{2}{c}{0.101 (0.076)} \\
Wtd.\ Med.  &  \multicolumn{2}{c}{0.112 (0.094)} &  \multicolumn{2}{c}{0.11 (0.107)} &  \multicolumn{2}{c}{0.205 (0.087)} \\
\hline  & MLE & Shrinkage & MLE & Shrinkage & MLE & \multicolumn{1}{c}{Shrinkage} \\
$\tau_2 = 0$,$\psi_I$  & 0.149 (0.05) & 0.157 (0.049) & 0.157 (0.071) & 0.156 (0.071) & 0.144 (0.069) & 0.154 (0.068) \\
$\tau_2 = 0$,$\psi_H$  & 0.166 (0.051) & 0.177 (0.05) & 0.176 (0.073) & 0.175 (0.073) & 0.158 (0.07) & 0.175 (0.069) \\
$\tau_2 \ne 0$,$\psi_I$  & 0.148 (0.051) & 0.156 (0.051) & 0.157 (0.083) & 0.157 (0.083) & 0.142 (0.07) & 0.152 (0.069) \\
$\tau_2 \ne 0$,$\psi_H$  & 0.163 (0.053) & 0.174 (0.052) & 0.177 (0.082) & 0.176 (0.082) & 0.153 (0.072) & 0.169 (0.071) \\
\hline
\end{tabular}
\caption{Screening: BMI (Dataset: Akiyama et al. (2017)); Exposure:
  BMI (Dataset: UK BioBank); Outcome: IS (Dataset: Malik et al. (2018)).}
\end{subtable}
\caption{Comparison of different MR methods to estimate the effects of
  Body Mass Index (BMI) on Coronary Artery Disease (CAD) and ischemic
  stroke (IS).}
\label{tab:bmi-detailed}
\end{table}

\begin{table} \centering \small
\begin{subtable}[t]{\textwidth} \centering
\begin{tabular}{l|cccccc}
\hline
 & \multicolumn{2}{c}{$p_{\mathrm{sel}} \in (0, 1)$} & \multicolumn{2}{c}{$p_{\mathrm{sel}} \in (0, 5 \times 10^{-8})$} & \multicolumn{2}{c}{$p_{\mathrm{sel}} \in (5 \times 10^{-8}, 1)$} \\

\hline
\# SNPs  &  \multicolumn{2}{c}{1214} &  \multicolumn{2}{c}{37} &  \multicolumn{2}{c}{1177} \\
$p_1$  &  \multicolumn{2}{c}{0.92} &  \multicolumn{2}{c}{0.99} &  \multicolumn{2}{c}{0.85} \\
$\sigma_1$  &  \multicolumn{2}{c}{0.54} &  \multicolumn{2}{c}{10.87} &  \multicolumn{2}{c}{0.34} \\
$\sigma_2$  &  \multicolumn{2}{c}{7.32} &  \multicolumn{2}{c}{10.87} &  \multicolumn{2}{c}{2.08} \\
MR-Egger  &  \multicolumn{2}{c}{0.3909 (0.0316)} &  \multicolumn{2}{c}{0.5047 (0.0877)} &  \multicolumn{2}{c}{0.0565 (0.0644)} \\
Wtd.\ Med.  &  \multicolumn{2}{c}{0.4215 (0.0486)} &  \multicolumn{2}{c}{0.4872 (0.0476)} &  \multicolumn{2}{c}{0.1825 (0.0699)} \\
\hline  & MLE & Shrinkage & MLE & Shrinkage & MLE & \multicolumn{1}{c}{Shrinkage} \\
$\tau_2 = 0$,$\psi_I$  & 0.475 (0.031) & 0.479 (0.029) & 0.473 (0.029) & 0.473 (0.029) & 0.489 (0.164) & 0.535 (0.134) \\
$\tau_2 = 0$,$\psi_H$  & 0.499 (0.031) & 0.497 (0.028) & 0.49 (0.029) & 0.49 (0.029) & 0.562 (0.126) & 0.567 (0.106) \\
$\tau_2 \ne 0$,$\psi_I$  & 0.424 (0.033) & 0.449 (0.032) & 0.445 (0.047) & 0.445 (0.047) & 0.259 (0.109) & 0.329 (0.116) \\
$\tau_2 \ne 0$,$\psi_H$  & 0.458 (0.033) & 0.477 (0.031) & 0.476 (0.043) & 0.476 (0.043) & 0.345 (0.11) & 0.42 (0.102) \\
\hline
\end{tabular}
\caption{Screening: LDL-C (Dataset: Teslovich et al. (2010)); Exposure:
  LDL-C (Dataset: Metabochip, Willer et al. (2013)); Outcome: CAD, (Dataset: CARDIoGRAMplusC4D).}
\end{subtable}
\begin{subtable}[t]{\textwidth} \centering
\begin{tabular}{l|cccccc}
\hline
 & \multicolumn{2}{c}{$p_{\mathrm{sel}} \in (0, 1)$} & \multicolumn{2}{c}{$p_{\mathrm{sel}} \in (0, 5 \times 10^{-8})$} & \multicolumn{2}{c}{$p_{\mathrm{sel}} \in (5 \times 10^{-8}, 1)$} \\

\hline
\# SNPs  &  \multicolumn{2}{c}{1191} &  \multicolumn{2}{c}{42} &  \multicolumn{2}{c}{1149} \\
$p_1$  &  \multicolumn{2}{c}{0.89} &  \multicolumn{2}{c}{0.01} &  \multicolumn{2}{c}{0.86} \\
$\sigma_1$  &  \multicolumn{2}{c}{0.72} &  \multicolumn{2}{c}{8.94} &  \multicolumn{2}{c}{0.61} \\
$\sigma_2$  &  \multicolumn{2}{c}{5.75} &  \multicolumn{2}{c}{8.94} &  \multicolumn{2}{c}{2.82} \\
MR-Egger  &  \multicolumn{2}{c}{-0.172 (0.0364)} &  \multicolumn{2}{c}{0.1911 (0.1138)} &  \multicolumn{2}{c}{-0.2896 (0.0597)} \\
Wtd.\ Med.  &  \multicolumn{2}{c}{0.0111 (0.056)} &  \multicolumn{2}{c}{0.0461 (0.0691)} &  \multicolumn{2}{c}{-0.0824 (0.0615)} \\
\hline  & MLE & Shrinkage & MLE & Shrinkage & MLE & \multicolumn{1}{c}{Shrinkage} \\
$\tau_2 = 0$,$\psi_I$  & -0.25 (0.034) & -0.212 (0.032) & -0.151 (0.035) & -0.151 (0.035) & -0.479 (0.078) & -0.421 (0.07) \\
$\tau_2 = 0$,$\psi_H$  & -0.325 (0.033) & -0.275 (0.031) & -0.238 (0.035) & -0.238 (0.035) & -0.431 (0.071) & -0.349 (0.065) \\
$\tau_2 \ne 0$,$\psi_I$  & -0.229 (0.037) & -0.201 (0.036) & -0.171 (0.069) & -0.171 (0.069) & -0.383 (0.073) & -0.35 (0.069) \\
$\tau_2 \ne 0$,$\psi_H$  & -0.282 (0.036) & -0.245 (0.035) & -0.201 (0.07) & -0.201 (0.07) & -0.353 (0.072) & -0.294 (0.067) \\
\hline
\end{tabular}
\caption{Screening: HDL-C (Dataset: Teslovich et al. (2010)); Exposure:
  HDL-C (Dataset: Metabochip, Willer et al. (2013)); Outcome: CAD, (Dataset: CARDIoGRAMplusC4D).}
\end{subtable}
\begin{subtable}[t]{\textwidth} \centering
\begin{tabular}{l|cccccc}
\hline
 & \multicolumn{2}{c}{$p_{\mathrm{sel}} \in (0, 1)$} & \multicolumn{2}{c}{$p_{\mathrm{sel}} \in (0, 5 \times 10^{-8})$} & \multicolumn{2}{c}{$p_{\mathrm{sel}} \in (5 \times 10^{-8}, 1)$} \\

\hline
\# SNPs  &  \multicolumn{2}{c}{1194} &  \multicolumn{2}{c}{28} &  \multicolumn{2}{c}{1166} \\
$p_1$  &  \multicolumn{2}{c}{0.95} &  \multicolumn{2}{c}{0.01} &  \multicolumn{2}{c}{0.83} \\
$\sigma_1$  &  \multicolumn{2}{c}{0.61} &  \multicolumn{2}{c}{11.48} &  \multicolumn{2}{c}{0.27} \\
$\sigma_2$  &  \multicolumn{2}{c}{8.21} &  \multicolumn{2}{c}{11.48} &  \multicolumn{2}{c}{1.93} \\
MR-Egger  &  \multicolumn{2}{c}{0.2037 (0.0355)} &  \multicolumn{2}{c}{0.0881 (0.0865)} &  \multicolumn{2}{c}{0.2767 (0.0761)} \\
Wtd.\ Med.  &  \multicolumn{2}{c}{0.1784 (0.054)} &  \multicolumn{2}{c}{0.1761 (0.0511)} &  \multicolumn{2}{c}{0.219 (0.0815)} \\
\hline  & MLE & Shrinkage & MLE & Shrinkage & MLE & \multicolumn{1}{c}{Shrinkage} \\
$\tau_2 = 0$,$\psi_I$  & 0.313 (0.036) & 0.261 (0.032) & 0.202 (0.032) & 0.202 (0.032) & 1.179 (0.19) & 1.039 (0.181) \\
$\tau_2 = 0$,$\psi_H$  & 0.344 (0.035) & 0.268 (0.032) & 0.213 (0.033) & 0.213 (0.033) & 1.091 (0.156) & 0.987 (0.134) \\
$\tau_2 \ne 0$,$\psi_I$  & 0.277 (0.038) & 0.248 (0.036) & 0.217 (0.051) & 0.217 (0.051) & 0.7 (0.144) & 0.709 (0.127) \\
$\tau_2 \ne 0$,$\psi_H$  & 0.288 (0.038) & 0.255 (0.036) & 0.217 (0.055) & 0.217 (0.055) & 0.726 (0.145) & 0.735 (0.125) \\
\hline
\end{tabular}
\caption{Screening: TG (Dataset: Teslovich et al. (2010)); Exposure:
  TG (Dataset: Metabochip, Willer et al. (2013)); Outcome: CAD, (Dataset: CARDIoGRAMplusC4D).}
\end{subtable}
\caption{Comparison of different MR methods to estimate the effects of blood lipid levels on Coronary Artery Disease (CAD) using the CARDIoGRAMplusC4D dataset and unrestricted instruments.}
\label{tab:lipid1-detailed}
\end{table}

\begin{table} \centering \small
\begin{subtable}[t]{\textwidth}\centering
\begin{tabular}{l|cccccc}
\hline
 & \multicolumn{2}{c}{$p_{\mathrm{sel}} \in (0, 1)$} & \multicolumn{2}{c}{$p_{\mathrm{sel}} \in (0, 5 \times 10^{-8})$} & \multicolumn{2}{c}{$p_{\mathrm{sel}} \in (5 \times 10^{-8}, 1)$} \\

\hline
\# SNPs  &  \multicolumn{2}{c}{898} &  \multicolumn{2}{c}{11} &  \multicolumn{2}{c}{887} \\
$p_1$  &  \multicolumn{2}{c}{0.92} &  \multicolumn{2}{c}{0.01} &  \multicolumn{2}{c}{0.92} \\
$\sigma_1$  &  \multicolumn{2}{c}{0.47} &  \multicolumn{2}{c}{5.93} &  \multicolumn{2}{c}{0.44} \\
$\sigma_2$  &  \multicolumn{2}{c}{3.38} &  \multicolumn{2}{c}{5.93} &  \multicolumn{2}{c}{2.39} \\
MR-Egger  &  \multicolumn{2}{c}{0.3167 (0.0614)} &  \multicolumn{2}{c}{0.7336 (0.3916)} &  \multicolumn{2}{c}{0.2563 (0.0731)} \\
Wtd.\ Med.  &  \multicolumn{2}{c}{0.3027 (0.0821)} &  \multicolumn{2}{c}{0.4638 (0.1241)} &  \multicolumn{2}{c}{0.2291 (0.0786)} \\
\hline  & MLE & Shrinkage & MLE & Shrinkage & MLE & \multicolumn{1}{c}{Shrinkage} \\
$\tau_2 = 0$,$\psi_I$  & 0.432 (0.089) & 0.504 (0.081) & 0.419 (0.101) & 0.419 (0.101) & 0.443 (0.135) & 0.577 (0.13) \\
$\tau_2 = 0$,$\psi_H$  & 0.364 (0.085) & 0.483 (0.076) & 0.44 (0.098) & 0.44 (0.098) & 0.29 (0.126) & 0.514 (0.116) \\
$\tau_2 \ne 0$,$\psi_I$  & 0.393 (0.086) & 0.482 (0.08) & 0.449 (0.167) & 0.449 (0.167) & 0.395 (0.128) & 0.538 (0.123) \\
$\tau_2 \ne 0$,$\psi_H$  & 0.348 (0.085) & 0.471 (0.077) & 0.452 (0.167) & 0.452 (0.167) & 0.277 (0.126) & 0.498 (0.116) \\
\hline
\end{tabular}
\caption{Screening: LDL-C (Dataset: Teslovich et al. (2010)); Exposure:
  LDL-C (Dataset: Metabochip, Willer et al. (2013)); Outcome: CAD,
  (Dataset: CARDIoGRAMplusC4D).}
\label{tab:lipid3-ldl}
\end{subtable}
\begin{subtable}[t]{\textwidth}\centering
\begin{tabular}{l|cccccc}
\hline
 & \multicolumn{2}{c}{$p_{\mathrm{sel}} \in (0, 1)$} & \multicolumn{2}{c}{$p_{\mathrm{sel}} \in (0, 5 \times 10^{-8})$} & \multicolumn{2}{c}{$p_{\mathrm{sel}} \in (5 \times 10^{-8}, 1)$} \\

\hline
\# SNPs  &  \multicolumn{2}{c}{869} &  \multicolumn{2}{c}{8} &  \multicolumn{2}{c}{861} \\
$p_1$  &  \multicolumn{2}{c}{0.95} &  \multicolumn{2}{c}{0.01} &  \multicolumn{2}{c}{0.93} \\
$\sigma_1$  &  \multicolumn{2}{c}{0.68} &  \multicolumn{2}{c}{7.07} &  \multicolumn{2}{c}{0.65} \\
$\sigma_2$  &  \multicolumn{2}{c}{3.88} &  \multicolumn{2}{c}{7.07} &  \multicolumn{2}{c}{2.38} \\
MR-Egger  &  \multicolumn{2}{c}{-0.0757 (0.0704)} &  \multicolumn{2}{c}{-0.1905 (0.3977)} &  \multicolumn{2}{c}{-0.057 (0.0839)} \\
Wtd.\ Med.  &  \multicolumn{2}{c}{-0.1496 (0.0812)} &  \multicolumn{2}{c}{-0.0792 (0.1235)} &  \multicolumn{2}{c}{-0.1836 (0.0861)} \\
\hline  & MLE & Shrinkage & MLE & Shrinkage & MLE & \multicolumn{1}{c}{Shrinkage} \\
$\tau_2 = 0$,$\psi_I$  & -0.402 (0.107) & -0.377 (0.086) & -0.142 (0.097) & -0.142 (0.097) & -0.622 (0.182) & -0.676 (0.155) \\
$\tau_2 = 0$,$\psi_H$  & -0.419 (0.094) & -0.38 (0.081) & -0.139 (0.1) & -0.139 (0.1) & -0.651 (0.143) & -0.707 (0.13) \\
$\tau_2 \ne 0$,$\psi_I$  & -0.276 (0.093) & -0.296 (0.085) & -0.142 (0.097) & -0.142 (0.097) & -0.348 (0.13) & -0.43 (0.124) \\
$\tau_2 \ne 0$,$\psi_H$  & -0.318 (0.092) & -0.318 (0.085) & -0.139 (0.1) & -0.139 (0.1) & -0.447 (0.131) & -0.53 (0.125) \\
\hline
\end{tabular}
\caption{Screening: HDL-C (Dataset: Teslovich et al. (2010)); Exposure:
  HDL-C (Dataset: Metabochip, Willer et al. (2013)); Outcome: CAD,
  (Dataset: CARDIoGRAMplusC4D).}
\label{tab:lipid3-hdl}
\end{subtable}
\begin{subtable}[t]{\textwidth}\centering
\begin{tabular}{l|cccccc}
\hline
 & \multicolumn{2}{c}{$p_{\mathrm{sel}} \in (0, 1)$} & \multicolumn{2}{c}{$p_{\mathrm{sel}} \in (0, 5 \times 10^{-8})$} & \multicolumn{2}{c}{$p_{\mathrm{sel}} \in (5 \times 10^{-8}, 1)$} \\

\hline
\# SNPs  &  \multicolumn{2}{c}{881} &  \multicolumn{2}{c}{2} &  \multicolumn{2}{c}{879} \\
$p_1$  &  \multicolumn{2}{c}{0.99} &  \multicolumn{2}{c}{0.01} &  \multicolumn{2}{c}{0.98} \\
$\sigma_1$  &  \multicolumn{2}{c}{0.4} &  \multicolumn{2}{c}{8.3} &  \multicolumn{2}{c}{0.38} \\
$\sigma_2$  &  \multicolumn{2}{c}{4.58} &  \multicolumn{2}{c}{8.3} &  \multicolumn{2}{c}{2.49} \\
MR-Egger  &  \multicolumn{2}{c}{0.0787 (0.0822)} &  \multicolumn{2}{c}{-0.1803 (0.1427)} &  \multicolumn{2}{c}{0.1616 (0.0935)} \\
Wtd.\ Med.  &  \multicolumn{2}{c}{-0.0195 (0.1264)} &  \multicolumn{2}{c}{-0.1803 (0.1427)} &  \multicolumn{2}{c}{0.1299 (0.0948)} \\
\hline  & MLE & Shrinkage & MLE & Shrinkage & MLE & \multicolumn{1}{c}{Shrinkage} \\
$\tau_2 = 0$,$\psi_I$  & 0.207 (0.259) & 0.046 (0.136) & -0.182 (0.144) & -0.182 (0.144) & 0.886 (0.657) & 0.852 (0.374) \\
$\tau_2 = 0$,$\psi_H$  & 0.25 (0.225) & -0.064 (0.132) & -0.182 (0.148) & -0.182 (0.148) & 0.77 (0.46) & 0.972 (0.333) \\
$\tau_2 \ne 0$,$\psi_I$  & 0.142 (0.187) & 0.044 (0.136) & -0.182 (0.144) & -0.182 (0.144) & 0.381 (0.307) & 0.555 (0.255) \\
$\tau_2 \ne 0$,$\psi_H$  & 0.159 (0.192) & -0.047 (0.136) & -0.182 (0.148) & -0.182 (0.148) & 0.385 (0.315) & 0.464 (0.256) \\
\hline
\end{tabular}
\caption{Screening: TG (Dataset: Teslovich et al. (2010)); Exposure:
  TG (Dataset: Metabochip, Willer et al. (2013)); Outcome: CAD, (Dataset: CARDIoGRAMplusC4D).}
\end{subtable}
\caption{Comparison of different MR methods to estimate the effects of blood lipid levels on Coronary Artery Disease (CAD) using the CARDIoGRAMplusC4D dataset and restricted instruments.}
\label{tab:lipid3-detailed}
\end{table}

\begin{table} \centering \small
\begin{subtable}[t]{\textwidth}\centering
\begin{tabular}{l|cccccc}
\hline
 & \multicolumn{2}{c}{$p_{\mathrm{sel}} \in (0, 1)$} & \multicolumn{2}{c}{$p_{\mathrm{sel}} \in (0, 5 \times 10^{-8})$} & \multicolumn{2}{c}{$p_{\mathrm{sel}} \in (5 \times 10^{-8}, 1)$} \\

\hline
\# SNPs  &  \multicolumn{2}{c}{1140} &  \multicolumn{2}{c}{37} &  \multicolumn{2}{c}{1103} \\
$p_1$  &  \multicolumn{2}{c}{0.92} &  \multicolumn{2}{c}{0.99} &  \multicolumn{2}{c}{0.87} \\
$\sigma_1$  &  \multicolumn{2}{c}{0.55} &  \multicolumn{2}{c}{10.87} &  \multicolumn{2}{c}{0.4} \\
$\sigma_2$  &  \multicolumn{2}{c}{7.27} &  \multicolumn{2}{c}{10.87} &  \multicolumn{2}{c}{2.34} \\
MR-Egger  &  \multicolumn{2}{c}{0.011 (0.0011)} &  \multicolumn{2}{c}{0.0127 (0.0023)} &  \multicolumn{2}{c}{0.0047 (0.0024)} \\
Wtd.\ Med.  &  \multicolumn{2}{c}{0.0122 (0.0017)} &  \multicolumn{2}{c}{0.0124 (0.0016)} &  \multicolumn{2}{c}{0.0061 (0.0027)} \\
\hline  & MLE & Shrinkage & MLE & Shrinkage & MLE & \multicolumn{1}{c}{Shrinkage} \\
$\tau_2 = 0$,$\psi_I$  & 0.012 (0.001) & 0.012 (0.001) & 0.012 (0.001) & 0.012 (0.001) & 0.014 (0.005) & 0.013 (0.004) \\
$\tau_2 = 0$,$\psi_H$  & 0.013 (0.001) & 0.013 (0.001) & 0.013 (0.001) & 0.013 (0.001) & 0.015 (0.004) & 0.014 (0.003) \\
$\tau_2 \ne 0$,$\psi_I$  & 0.012 (0.001) & 0.012 (0.001) & 0.012 (0.001) & 0.012 (0.001) & 0.01 (0.004) & 0.01 (0.003) \\
$\tau_2 \ne 0$,$\psi_H$  & 0.012 (0.001) & 0.012 (0.001) & 0.013 (0.001) & 0.013 (0.001) & 0.01 (0.004) & 0.011 (0.003) \\
\hline
\end{tabular}
\caption{Screening: LDL-C (Dataset: Teslovich et al. (2010)); Exposure:
  LDL-C (Dataset: Metabochip, Willer et al. (2013)); Outcome: CAD, (Dataset: UK BioBank).}
\end{subtable}

\begin{subtable}[t]{\textwidth}\centering
\begin{tabular}{l|cccccc}
\hline
 & \multicolumn{2}{c}{$p_{\mathrm{sel}} \in (0, 1)$} & \multicolumn{2}{c}{$p_{\mathrm{sel}} \in (0, 5 \times 10^{-8})$} & \multicolumn{2}{c}{$p_{\mathrm{sel}} \in (5 \times 10^{-8}, 1)$} \\

\hline
\# SNPs  &  \multicolumn{2}{c}{1079} &  \multicolumn{2}{c}{39} &  \multicolumn{2}{c}{1040} \\
$p_1$  &  \multicolumn{2}{c}{0.89} &  \multicolumn{2}{c}{0.01} &  \multicolumn{2}{c}{0.85} \\
$\sigma_1$  &  \multicolumn{2}{c}{0.76} &  \multicolumn{2}{c}{9.15} &  \multicolumn{2}{c}{0.64} \\
$\sigma_2$  &  \multicolumn{2}{c}{5.9} &  \multicolumn{2}{c}{9.15} &  \multicolumn{2}{c}{2.81} \\
MR-Egger  &  \multicolumn{2}{c}{-0.0056 (0.0013)} &  \multicolumn{2}{c}{0.0094 (0.0043)} &  \multicolumn{2}{c}{-0.0112 (0.0022)} \\
Wtd.\ Med.  &  \multicolumn{2}{c}{-0.0018 (0.0019)} &  \multicolumn{2}{c}{-7e-04 (0.0025)} &  \multicolumn{2}{c}{-0.0044 (0.0024)} \\
\hline  & MLE & Shrinkage & MLE & Shrinkage & MLE & \multicolumn{1}{c}{Shrinkage} \\
$\tau_2 = 0$,$\psi_I$  & -0.007 (0.001) & -0.007 (0.001) & -0.004 (0.001) & -0.004 (0.001) & -0.014 (0.003) & -0.014 (0.002) \\
$\tau_2 = 0$,$\psi_H$  & -0.009 (0.001) & -0.008 (0.001) & -0.006 (0.001) & -0.006 (0.001) & -0.012 (0.003) & -0.011 (0.002) \\
$\tau_2 \ne 0$,$\psi_I$  & -0.007 (0.001) & -0.006 (0.001) & -0.004 (0.003) & -0.004 (0.003) & -0.013 (0.003) & -0.014 (0.002) \\
$\tau_2 \ne 0$,$\psi_H$  & -0.009 (0.001) & -0.008 (0.001) & -0.004 (0.003) & -0.004 (0.003) & -0.011 (0.003) & -0.011 (0.002) \\
\hline
\end{tabular}
\caption{Screening: HDL-C (Dataset: Teslovich et al. (2010)); Exposure:
  HDL-C (Dataset: Metabochip, Willer et al. (2013)); Outcome: CAD, (Dataset: UK BioBank).}
\end{subtable}
\begin{subtable}[t]{\textwidth}\centering
\begin{tabular}{l|cccccc}
\hline
 & \multicolumn{2}{c}{$p_{\mathrm{sel}} \in (0, 1)$} & \multicolumn{2}{c}{$p_{\mathrm{sel}} \in (0, 5 \times 10^{-8})$} & \multicolumn{2}{c}{$p_{\mathrm{sel}} \in (5 \times 10^{-8}, 1)$} \\

\hline
\# SNPs  &  \multicolumn{2}{c}{1122} &  \multicolumn{2}{c}{27} &  \multicolumn{2}{c}{1095} \\
$p_1$  &  \multicolumn{2}{c}{0.95} &  \multicolumn{2}{c}{0.01} &  \multicolumn{2}{c}{0.84} \\
$\sigma_1$  &  \multicolumn{2}{c}{0.64} &  \multicolumn{2}{c}{11.68} &  \multicolumn{2}{c}{0.34} \\
$\sigma_2$  &  \multicolumn{2}{c}{8.26} &  \multicolumn{2}{c}{11.68} &  \multicolumn{2}{c}{1.99} \\
MR-Egger  &  \multicolumn{2}{c}{0.0088 (0.0013)} &  \multicolumn{2}{c}{0.0032 (0.0035)} &  \multicolumn{2}{c}{0.0139 (0.0027)} \\
Wtd.\ Med.  &  \multicolumn{2}{c}{0.0072 (0.0019)} &  \multicolumn{2}{c}{0.0072 (0.0018)} &  \multicolumn{2}{c}{0.0087 (0.003)} \\
\hline  & MLE & Shrinkage & MLE & Shrinkage & MLE & \multicolumn{1}{c}{Shrinkage} \\
$\tau_2 = 0$,$\psi_I$  & 0.013 (0.001) & 0.011 (0.001) & 0.008 (0.001) & 0.008 (0.001) & 0.033 (0.005) & 0.03 (0.004) \\
$\tau_2 = 0$,$\psi_H$  & 0.013 (0.001) & 0.011 (0.001) & 0.008 (0.001) & 0.008 (0.001) & 0.03 (0.004) & 0.026 (0.004) \\
$\tau_2 \ne 0$,$\psi_I$  & 0.012 (0.001) & 0.01 (0.001) & 0.007 (0.002) & 0.007 (0.002) & 0.029 (0.005) & 0.027 (0.004) \\
$\tau_2 \ne 0$,$\psi_H$  & 0.012 (0.001) & 0.01 (0.001) & 0.007 (0.002) & 0.007 (0.002) & 0.027 (0.005) & 0.024 (0.004) \\
\hline
\end{tabular}
\caption{Screening: TG (Dataset: Teslovich et al. (2010)); Exposure:
  TG (Dataset: Metabochip, Willer et al. (2013)); Outcome: CAD, (Dataset: UK BioBank).}
\end{subtable}
\caption{Comparison of different MR methods to estimate the effects of
  blood lipid levels on Coronary Artery Disease (CAD) using the UK
  BioBank dataset and unrestricted instruments.}
\label{tab:lipid2-detailed}
\end{table}

\begin{table} \centering \small
\begin{subtable}[t]{\textwidth}\centering
\begin{tabular}{l|cccccc}
\hline
 & \multicolumn{2}{c}{$p_{\mathrm{sel}} \in (0, 1)$} & \multicolumn{2}{c}{$p_{\mathrm{sel}} \in (0, 5 \times 10^{-8})$} & \multicolumn{2}{c}{$p_{\mathrm{sel}} \in (5 \times 10^{-8}, 1)$} \\

\hline
\# SNPs  &  \multicolumn{2}{c}{835} &  \multicolumn{2}{c}{11} &  \multicolumn{2}{c}{824} \\
$p_1$  &  \multicolumn{2}{c}{0.92} &  \multicolumn{2}{c}{0.01} &  \multicolumn{2}{c}{0.92} \\
$\sigma_1$  &  \multicolumn{2}{c}{0.45} &  \multicolumn{2}{c}{5.93} &  \multicolumn{2}{c}{0.42} \\
$\sigma_2$  &  \multicolumn{2}{c}{3.41} &  \multicolumn{2}{c}{5.93} &  \multicolumn{2}{c}{2.41} \\
MR-Egger  &  \multicolumn{2}{c}{0.0108 (0.0025)} &  \multicolumn{2}{c}{0.0165 (0.0132)} &  \multicolumn{2}{c}{0.0095 (0.0029)} \\
Wtd.\ Med.  &  \multicolumn{2}{c}{0.006 (0.0033)} &  \multicolumn{2}{c}{0.0068 (0.0049)} &  \multicolumn{2}{c}{0.0067 (0.0033)} \\
\hline  & MLE & Shrinkage & MLE & Shrinkage & MLE & \multicolumn{1}{c}{Shrinkage} \\
$\tau_2 = 0$,$\psi_I$  & 0.015 (0.004) & 0.014 (0.003) & 0.013 (0.004) & 0.013 (0.004) & 0.018 (0.006) & 0.016 (0.006) \\
$\tau_2 = 0$,$\psi_H$  & 0.015 (0.004) & 0.013 (0.003) & 0.012 (0.004) & 0.012 (0.004) & 0.018 (0.006) & 0.017 (0.005) \\
$\tau_2 \ne 0$,$\psi_I$  & 0.014 (0.004) & 0.014 (0.003) & 0.014 (0.006) & 0.014 (0.006) & 0.015 (0.006) & 0.015 (0.005) \\
$\tau_2 \ne 0$,$\psi_H$  & 0.013 (0.004) & 0.013 (0.003) & 0.014 (0.006) & 0.014 (0.006) & 0.015 (0.005) & 0.015 (0.005) \\
\hline
\end{tabular}
\caption{Screening: LDL-C (Dataset: Teslovich et al. (2010)); Exposure:
  LDL-C (Dataset: Metabochip, Willer et al. (2013)); Outcome: CAD, (Dataset: UK BioBank).}
\end{subtable}
\begin{subtable}[t]{\textwidth}\centering
\begin{tabular}{l|cccccc}
\hline
 & \multicolumn{2}{c}{$p_{\mathrm{sel}} \in (0, 1)$} & \multicolumn{2}{c}{$p_{\mathrm{sel}} \in (0, 5 \times 10^{-8})$} & \multicolumn{2}{c}{$p_{\mathrm{sel}} \in (5 \times 10^{-8}, 1)$} \\

\hline
\# SNPs  &  \multicolumn{2}{c}{800} &  \multicolumn{2}{c}{7} &  \multicolumn{2}{c}{793} \\
$p_1$  &  \multicolumn{2}{c}{0.95} &  \multicolumn{2}{c}{0.01} &  \multicolumn{2}{c}{0.95} \\
$\sigma_1$  &  \multicolumn{2}{c}{0.68} &  \multicolumn{2}{c}{7.09} &  \multicolumn{2}{c}{0.66} \\
$\sigma_2$  &  \multicolumn{2}{c}{3.99} &  \multicolumn{2}{c}{7.09} &  \multicolumn{2}{c}{2.59} \\
MR-Egger  &  \multicolumn{2}{c}{-0.0054 (0.0025)} &  \multicolumn{2}{c}{0.0088 (0.0157)} &  \multicolumn{2}{c}{-0.0077 (0.003)} \\
Wtd.\ Med.  &  \multicolumn{2}{c}{-0.0065 (0.0032)} &  \multicolumn{2}{c}{-0.0055 (0.005)} &  \multicolumn{2}{c}{-0.0084 (0.0034)} \\
\hline  & MLE & Shrinkage & MLE & Shrinkage & MLE & \multicolumn{1}{c}{Shrinkage} \\
$\tau_2 = 0$,$\psi_I$  & -0.012 (0.003) & -0.011 (0.003) & -0.001 (0.004) & -0.001 (0.004) & -0.018 (0.005) & -0.021 (0.004) \\
$\tau_2 = 0$,$\psi_H$  & -0.015 (0.003) & -0.013 (0.003) & -0.003 (0.004) & -0.003 (0.004) & -0.022 (0.005) & -0.024 (0.004) \\
$\tau_2 \ne 0$,$\psi_I$  & -0.011 (0.003) & -0.011 (0.003) & -0.001 (0.004) & -0.001 (0.004) & -0.017 (0.005) & -0.02 (0.004) \\
$\tau_2 \ne 0$,$\psi_H$  & -0.014 (0.003) & -0.013 (0.003) & -0.003 (0.004) & -0.003 (0.004) & -0.02 (0.005) & -0.023 (0.005) \\
\hline
\end{tabular}
\caption{Screening: HDL-C (Dataset: Teslovich et al. (2010)); Exposure:
  HDL-C (Dataset: Metabochip, Willer et al. (2013)); Outcome: CAD, (Dataset: UK BioBank).}
\end{subtable}
\begin{subtable}[t]{\textwidth}\centering
\begin{tabular}{l|cccccc}
\hline
 & \multicolumn{2}{c}{$p_{\mathrm{sel}} \in (0, 1)$} & \multicolumn{2}{c}{$p_{\mathrm{sel}} \in (0, 5 \times 10^{-8})$} & \multicolumn{2}{c}{$p_{\mathrm{sel}} \in (5 \times 10^{-8}, 1)$} \\

\hline
\# SNPs  &  \multicolumn{2}{c}{801} &  \multicolumn{2}{c}{2} &  \multicolumn{2}{c}{799} \\
$p_1$  &  \multicolumn{2}{c}{0.99} &  \multicolumn{2}{c}{0.01} &  \multicolumn{2}{c}{0.95} \\
$\sigma_1$  &  \multicolumn{2}{c}{0.4} &  \multicolumn{2}{c}{8.3} &  \multicolumn{2}{c}{0.32} \\
$\sigma_2$  &  \multicolumn{2}{c}{4.4} &  \multicolumn{2}{c}{8.3} &  \multicolumn{2}{c}{1.61} \\
MR-Egger  &  \multicolumn{2}{c}{0.0059 (0.0033)} &  \multicolumn{2}{c}{0.0036 (0.0056)} &  \multicolumn{2}{c}{0.0068 (0.0038)} \\
Wtd.\ Med.  &  \multicolumn{2}{c}{0.0064 (0.0042)} &  \multicolumn{2}{c}{0.0036 (0.0056)} &  \multicolumn{2}{c}{0.0043 (0.0039)} \\
\hline  & MLE & Shrinkage & MLE & Shrinkage & MLE & \multicolumn{1}{c}{Shrinkage} \\
$\tau_2 = 0$,$\psi_I$  & 0.026 (0.009) & 0.007 (0.005) & 0.004 (0.006) & 0.004 (0.006) & 0.046 (0.016) & 0.025 (0.017) \\
$\tau_2 = 0$,$\psi_H$  & 0.032 (0.009) & 0.005 (0.005) & 0.004 (0.006) & 0.004 (0.006) & 0.049 (0.017) & 0.03 (0.015) \\
$\tau_2 \ne 0$,$\psi_I$  & 0.021 (0.007) & 0.006 (0.005) & 0.004 (0.006) & 0.004 (0.006) & 0.033 (0.013) & 0.018 (0.011) \\
$\tau_2 \ne 0$,$\psi_H$  & 0.019 (0.008) & 0.005 (0.005) & 0.004 (0.006) & 0.004 (0.006) & 0.03 (0.014) & 0.017 (0.012) \\
\hline
\end{tabular}
\caption{Screening: TG (Dataset: Teslovich et al. (2010)); Exposure:
  TG (Dataset: Metabochip, Willer et al. (2013)); Outcome: CAD, (Dataset: UK BioBank).}
\end{subtable}
\caption{Comparison of different MR methods to estimate the effects of blood lipid levels on Coronary Artery Disease (CAD) using the UK BioBank dataset and restricted instruments.}
\label{tab:lipid4-detailed}
\end{table}

\newpage
\subsection{Diagnostic plots}
\label{sec:diagnostic-plots}

\begin{figure}[h]
  \centering
  \begin{subfigure}[b]{\textwidth} \centering
    \includegraphics[width = 0.75\textwidth]{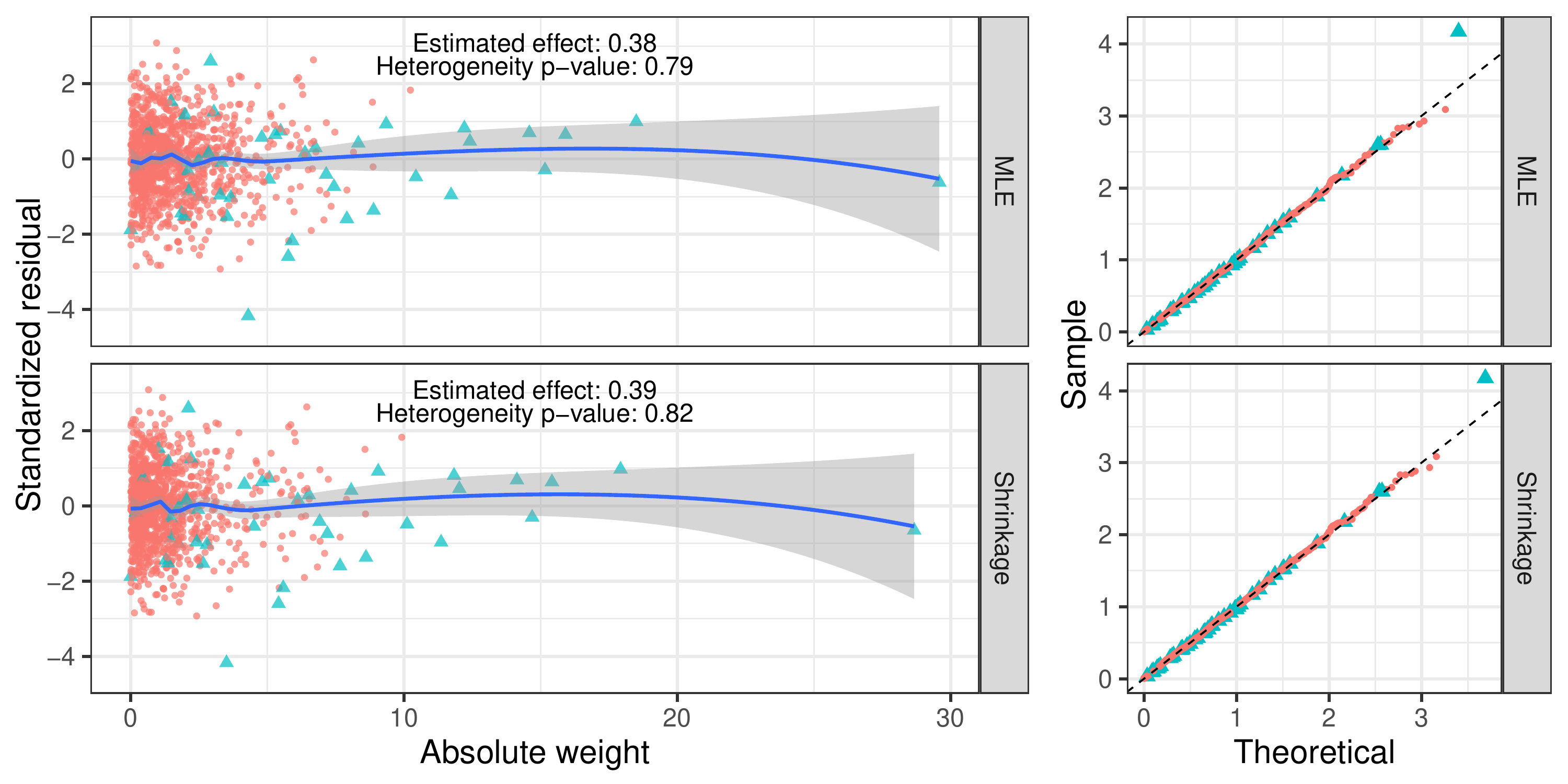}
    \caption{Screening: BMI (Dataset: Akiyama et al. (2017)); Exposure:
  BMI (Dataset: UK BioBank); Outcome: CAD, (Dataset: CARDIoGRAMplusC4D).}
  \end{subfigure}
  \begin{subfigure}[b]{\textwidth} \centering
    \includegraphics[width = 0.75\textwidth]{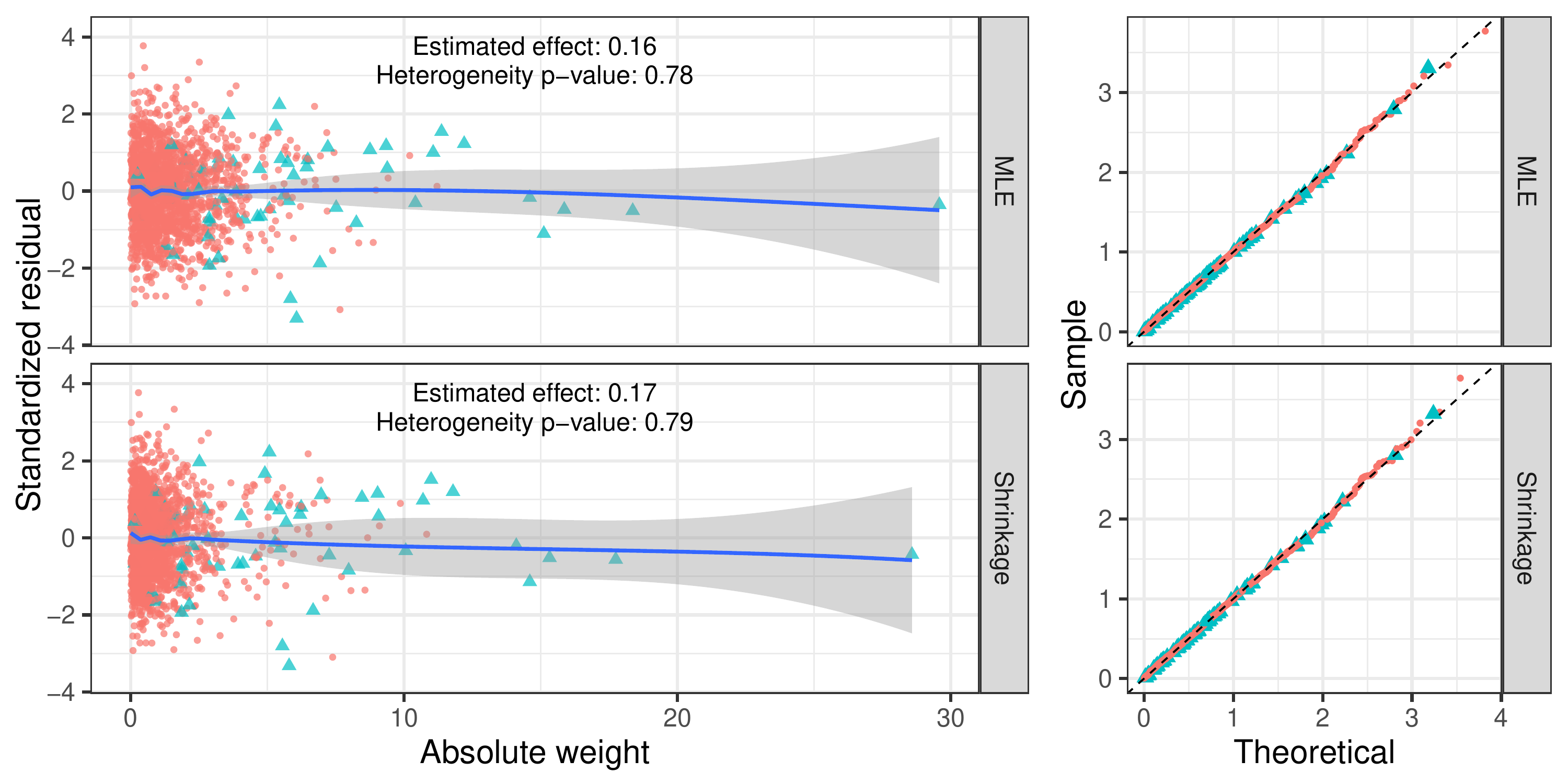}
    \caption{Screening: BMI (Dataset: Akiyama et al. (2017)); Exposure:
  BMI (Dataset: UK BioBank); Outcome: IS (Dataset: Malik et al. (2018)).}
  \end{subfigure}
  \caption{Diagnostic plots for the BMI results.}
  \label{fig:bmi-diagnostics}
\end{figure}

\begin{figure}
  \centering
  \begin{subfigure}[b]{\textwidth} \centering
    \includegraphics[width = 0.75\textwidth]{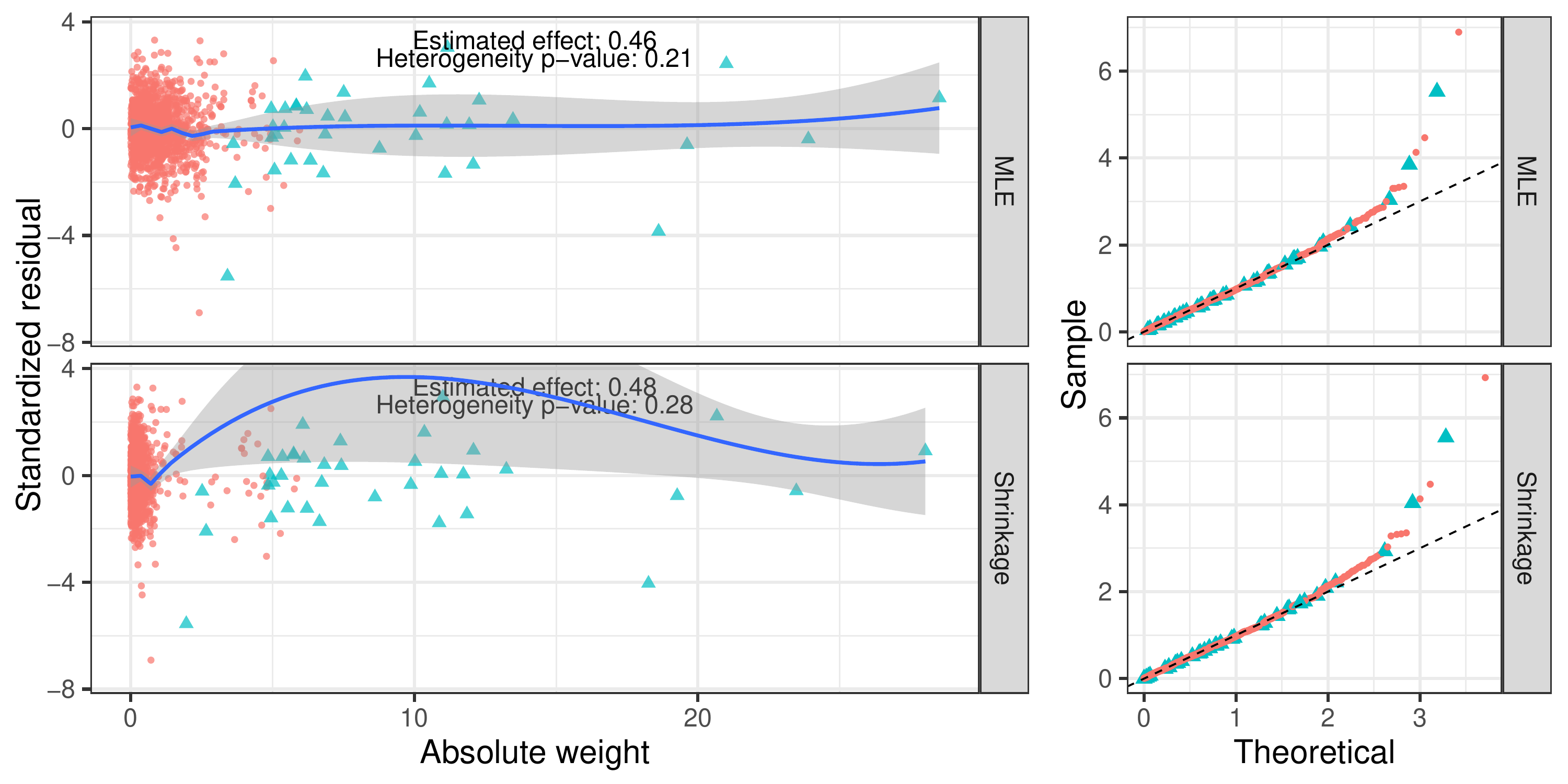}
    \caption{LDL-c.}
  \end{subfigure}
  \begin{subfigure}[b]{\textwidth} \centering
    \includegraphics[width = 0.75\textwidth]{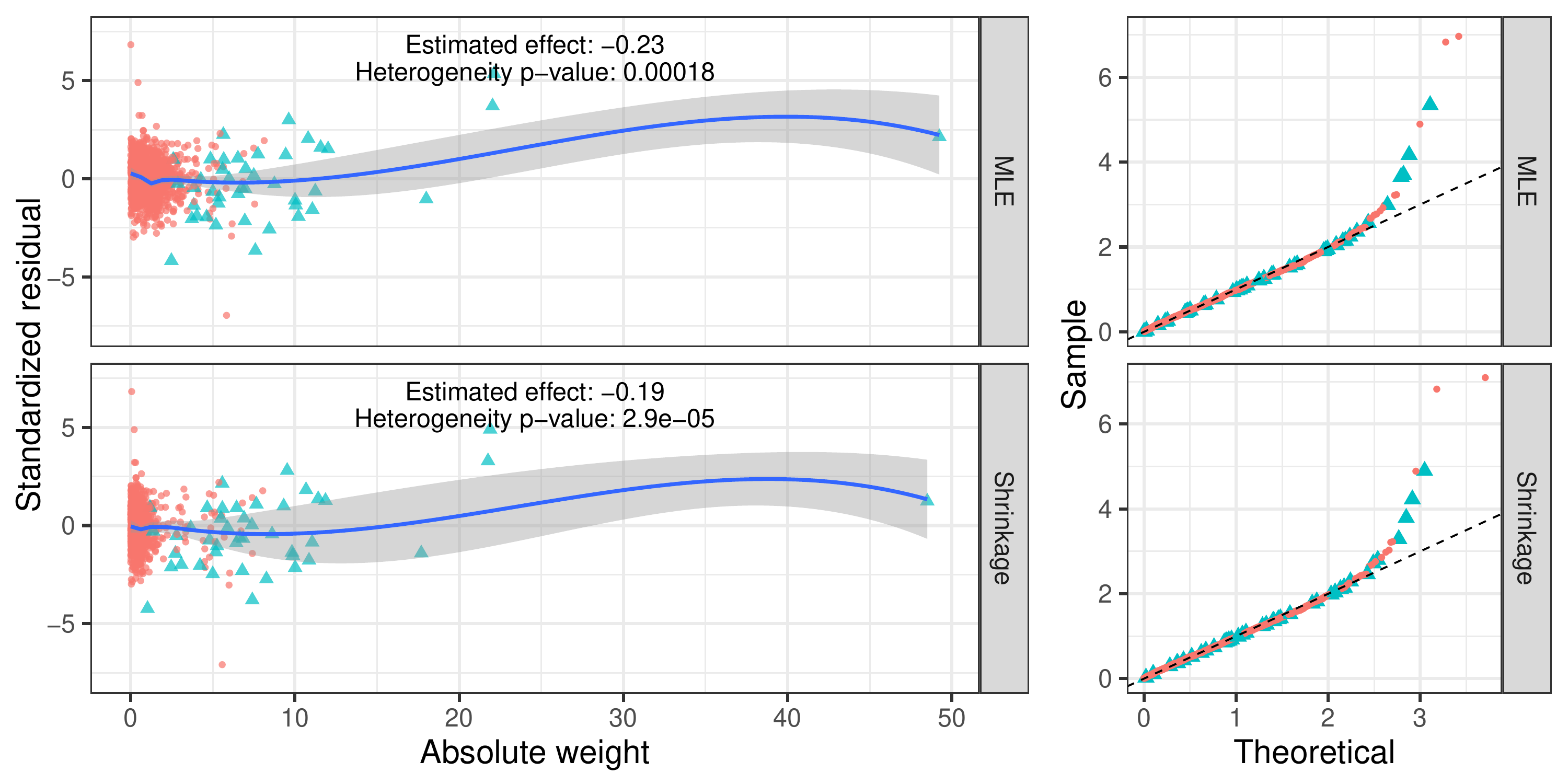}
    \caption{HDL-c.}
    \label{fig:hdc-diagnostics-1}
  \end{subfigure}
  \begin{subfigure}[b]{\textwidth} \centering
    \includegraphics[width = 0.75\textwidth]{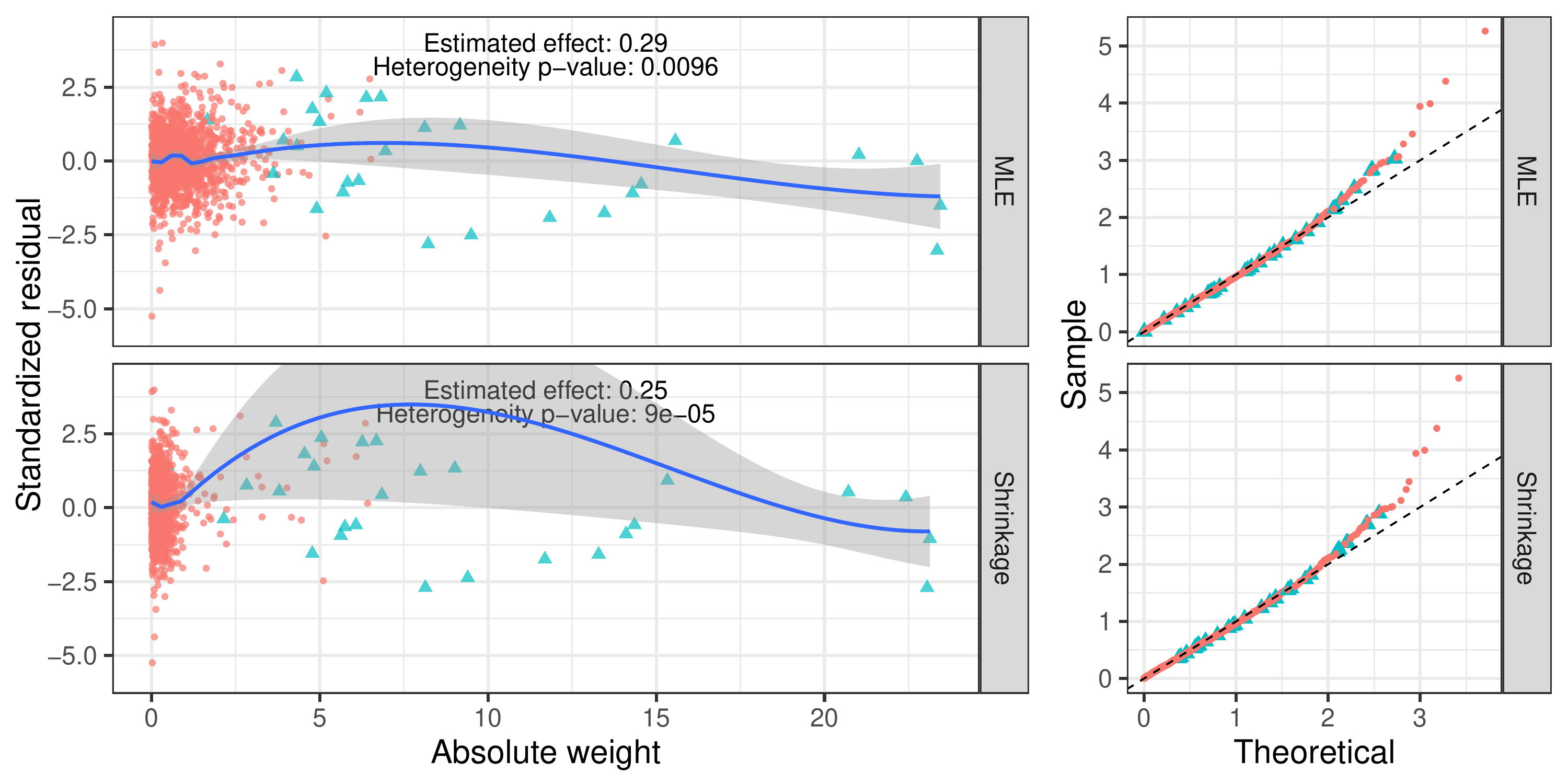}
    \caption{TG.}
  \end{subfigure}
  \caption{Diagnostic plots for the lipids results using unrestricted
    instruments and the CARDIoGRAMplusC4D dataset.}
  \label{fig:lipid-diagnostics-1}
\end{figure}

\begin{figure}
  \centering
  \begin{subfigure}[b]{\textwidth} \centering
    \includegraphics[width = 0.75\textwidth]{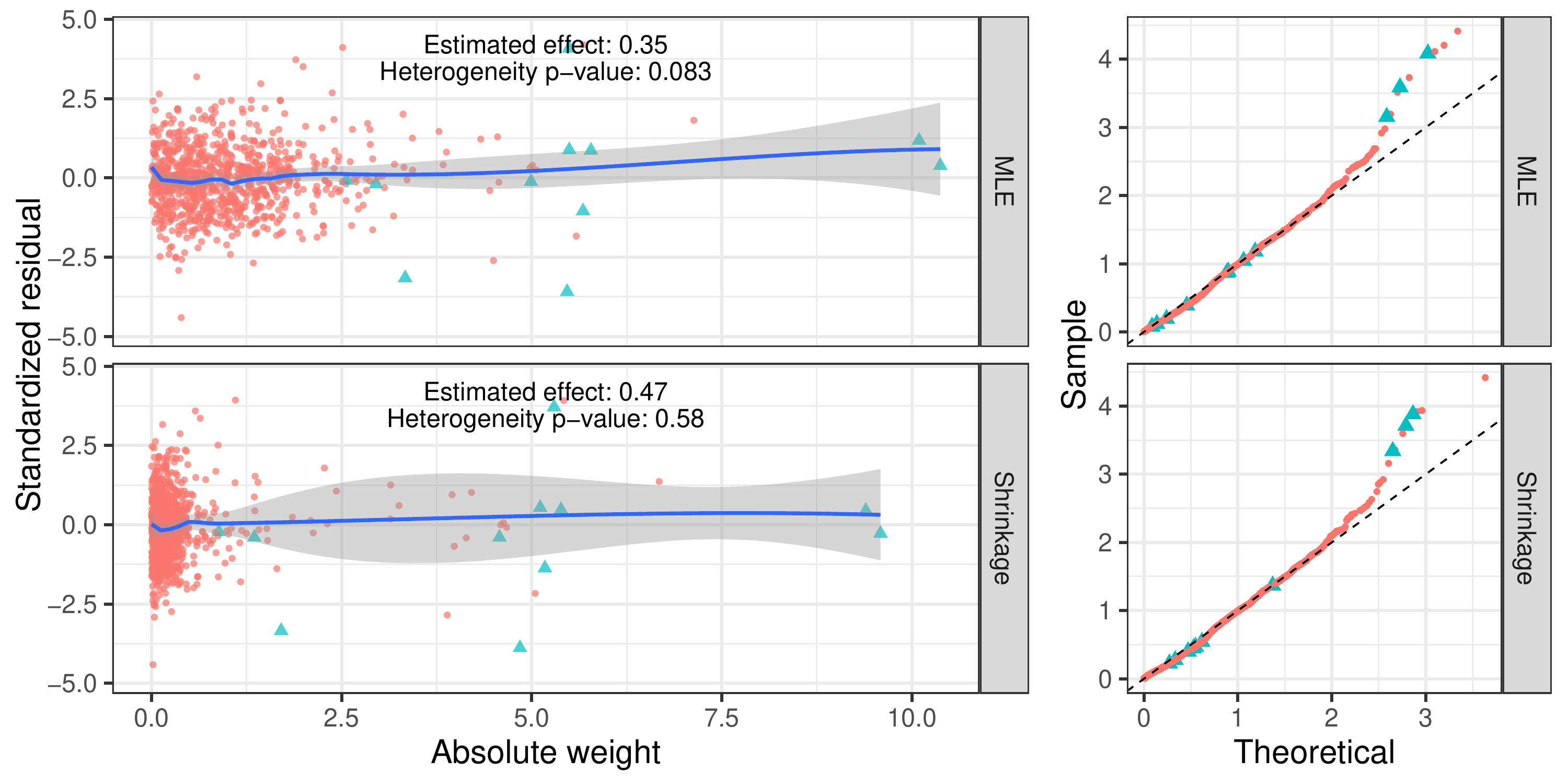}
    \caption{LDL-c.}
  \end{subfigure}
  \begin{subfigure}[b]{\textwidth} \centering
    \includegraphics[width = 0.75\textwidth]{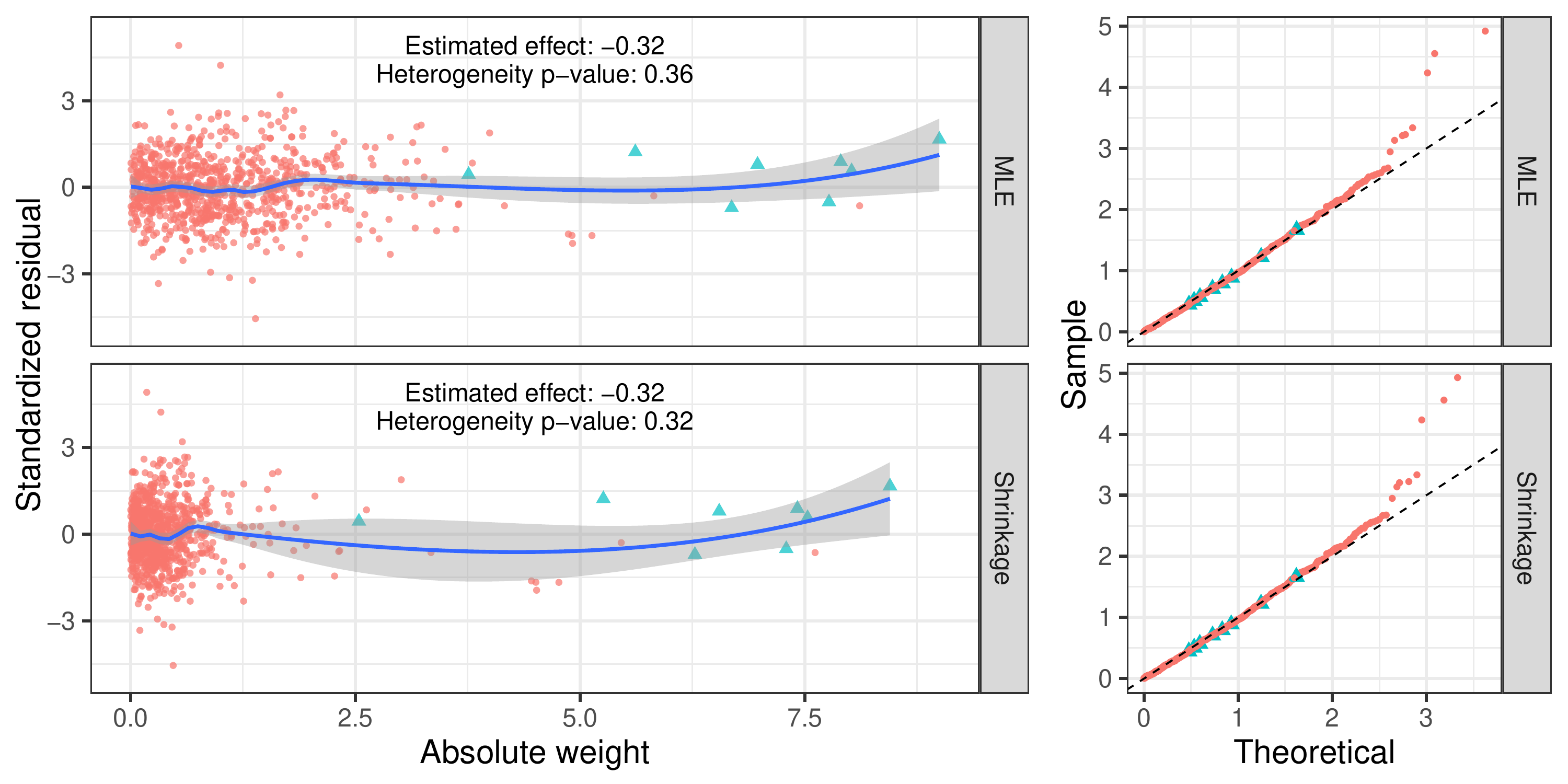}
    \caption{HDL-c.}
    \label{fig:hdc-diagnostics-2}
  \end{subfigure}
  \begin{subfigure}[b]{\textwidth} \centering
    \includegraphics[width = 0.75\textwidth]{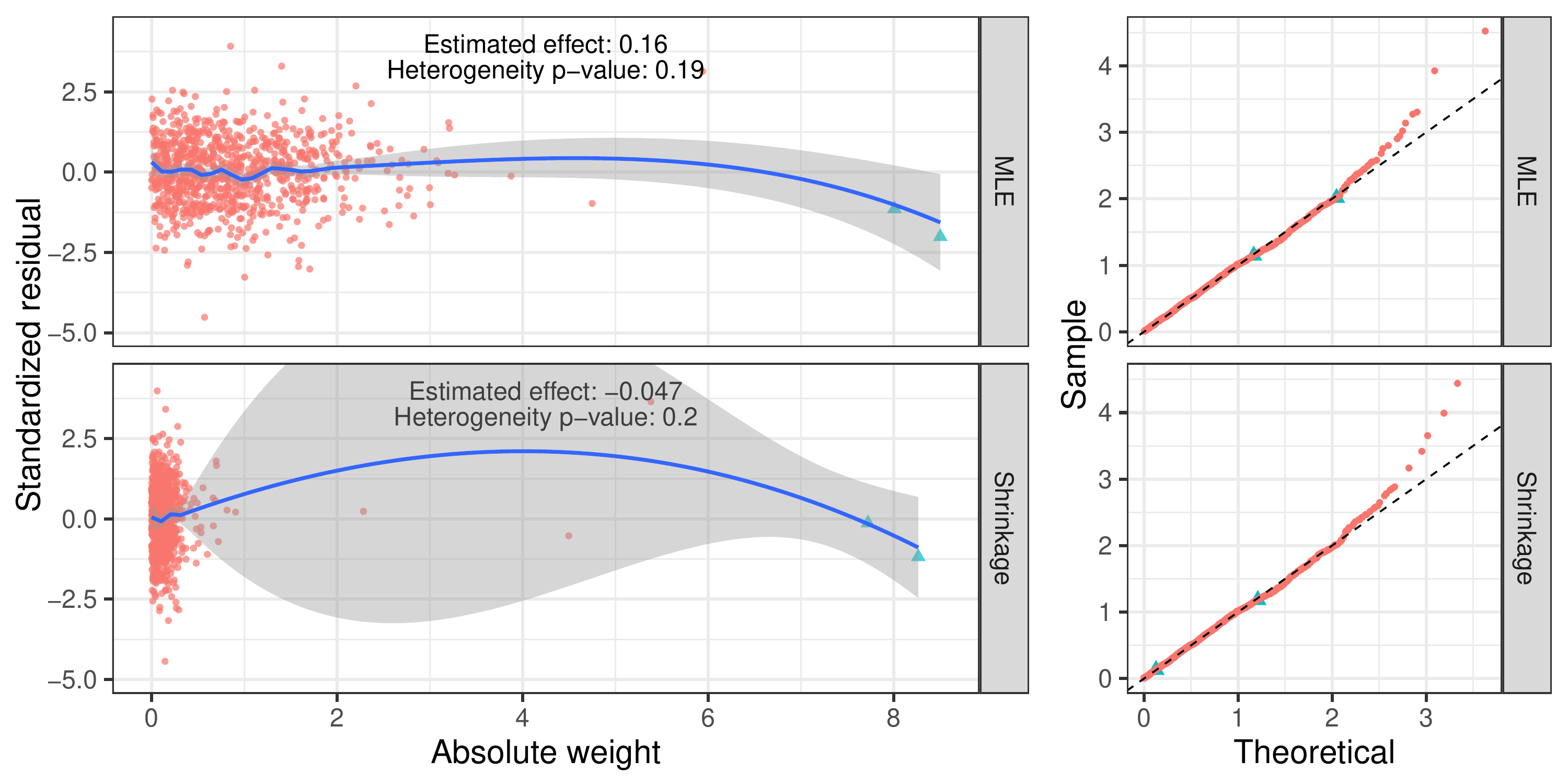}
    \caption{TG.}
  \end{subfigure}
  \caption{Diagnostic plots for the lipids results using restricted
    instruments and the CARDIoGRAMplusC4D dataset.}
  \label{fig:lipid-diagnostics-2}
\end{figure}

\begin{figure}
  \centering
  \begin{subfigure}[b]{\textwidth} \centering
    \includegraphics[width = 0.75\textwidth]{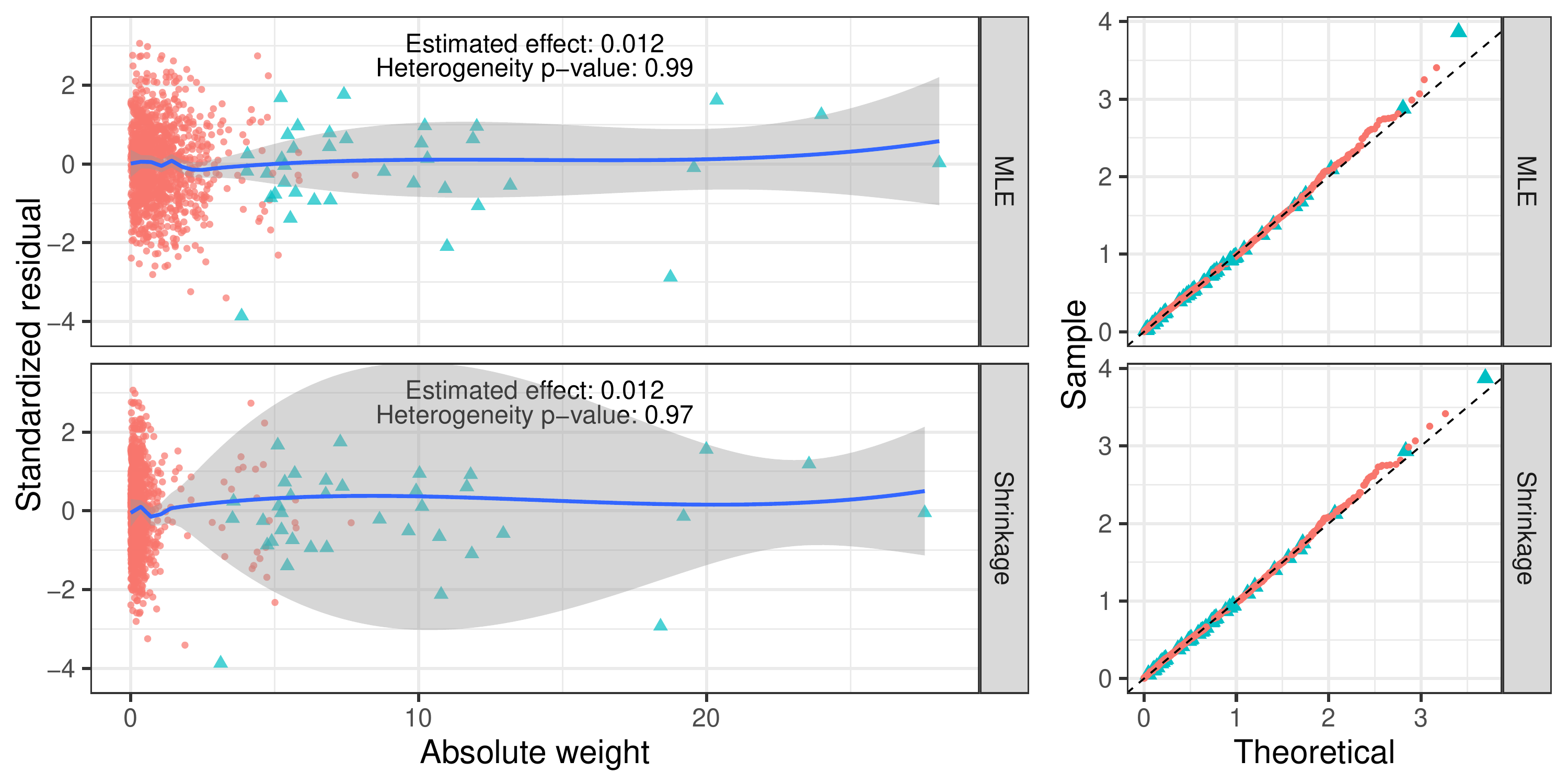}
    \caption{LDL-c.}
  \end{subfigure}
  \begin{subfigure}[b]{\textwidth} \centering
    \includegraphics[width = 0.75\textwidth]{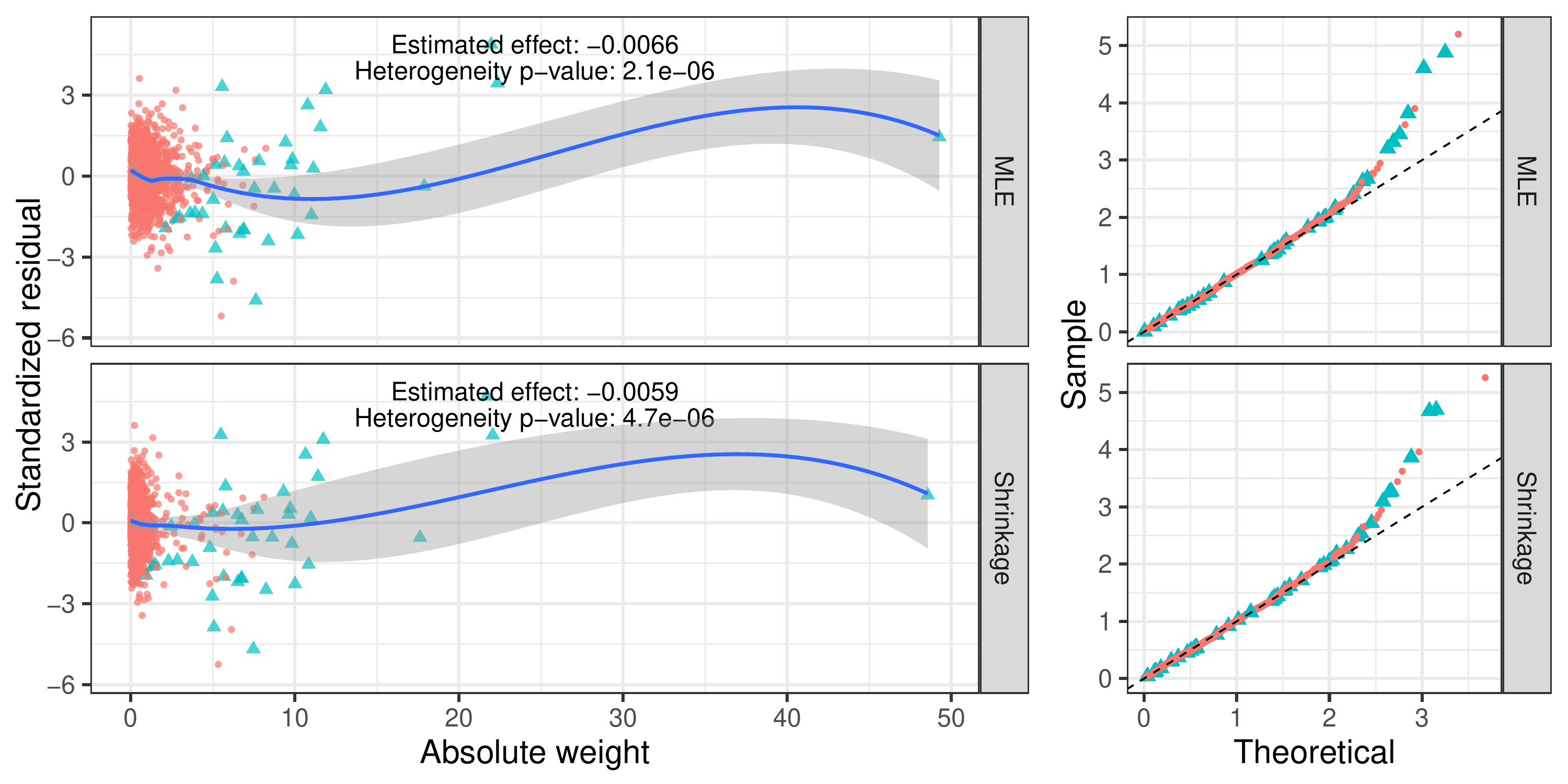}
    \caption{HDL-c.}
    \label{fig:hdc-diagnostics-3}
  \end{subfigure}
  \begin{subfigure}[b]{\textwidth} \centering
    \includegraphics[width = 0.75\textwidth]{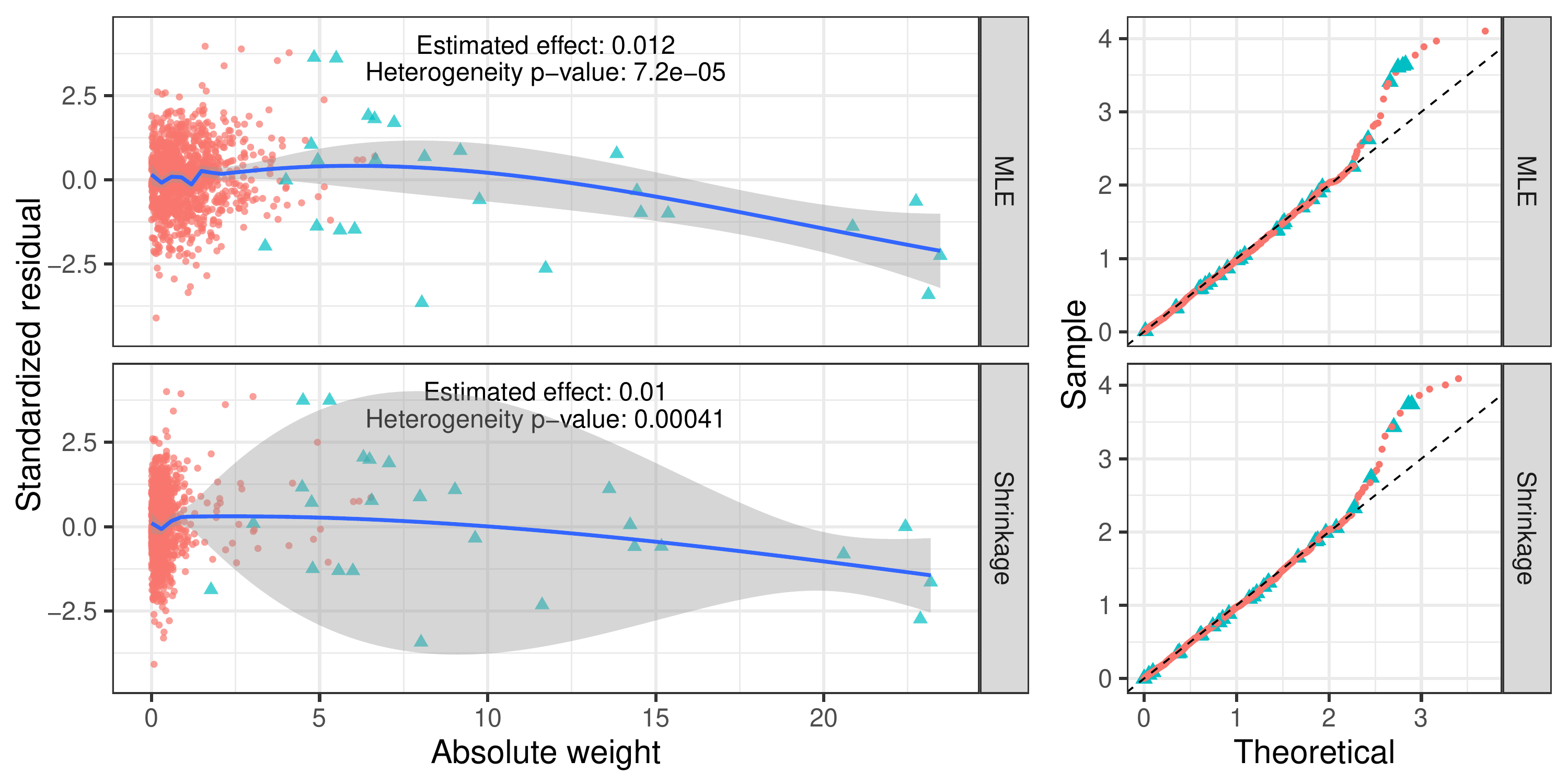}
    \caption{TG.}
  \end{subfigure}
  \caption{Diagnostic plots for the lipids results using unrestricted
    instruments and the UK BioBank dataset dataset.}
  \label{fig:lipid-diagnostics-3}
\end{figure}

\begin{figure}
  \centering
  \begin{subfigure}[b]{\textwidth} \centering
    \includegraphics[width = 0.75\textwidth]{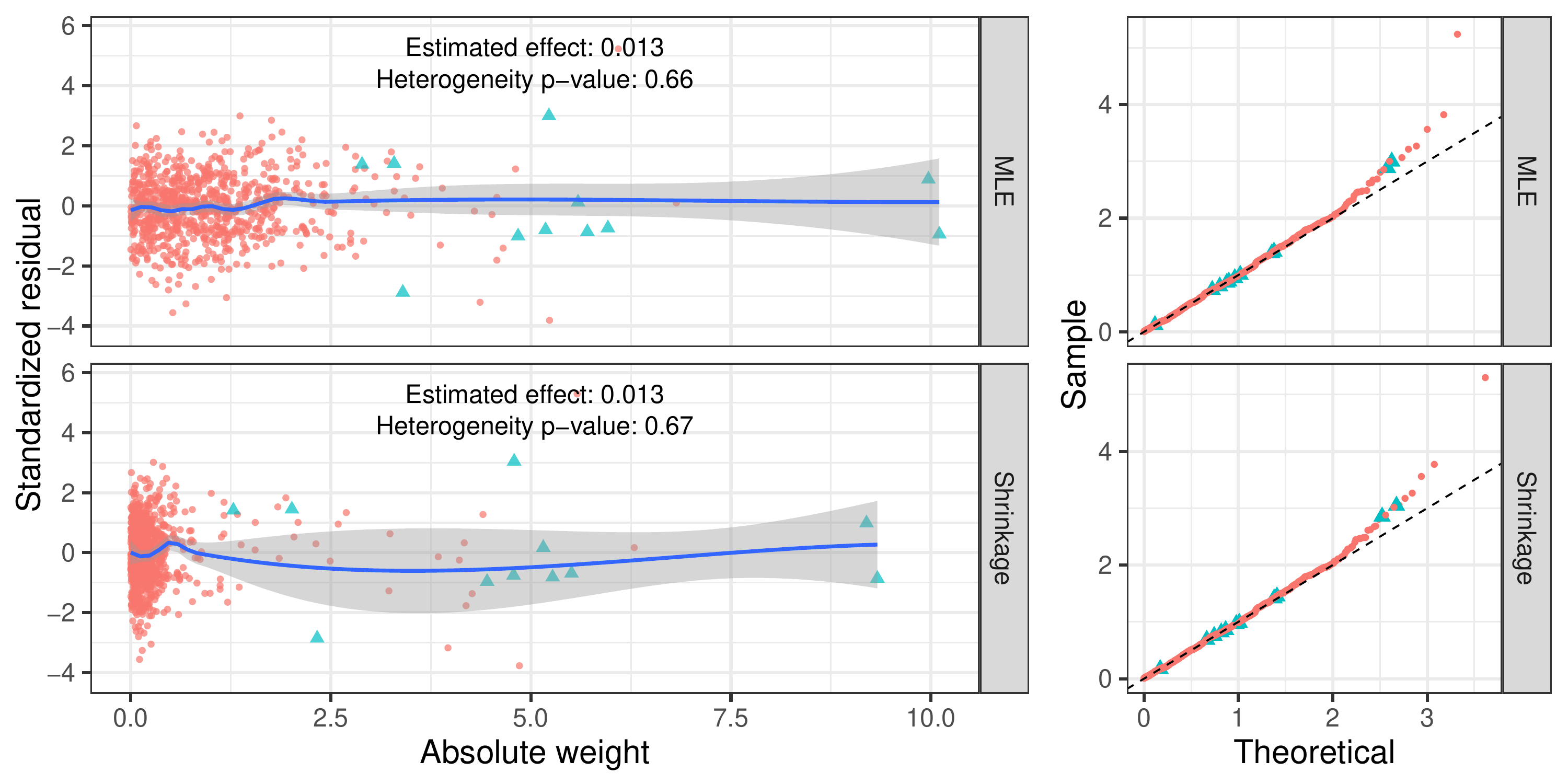}
    \caption{LDL-c.}
  \end{subfigure}
  \begin{subfigure}[b]{\textwidth} \centering
    \includegraphics[width = 0.75\textwidth]{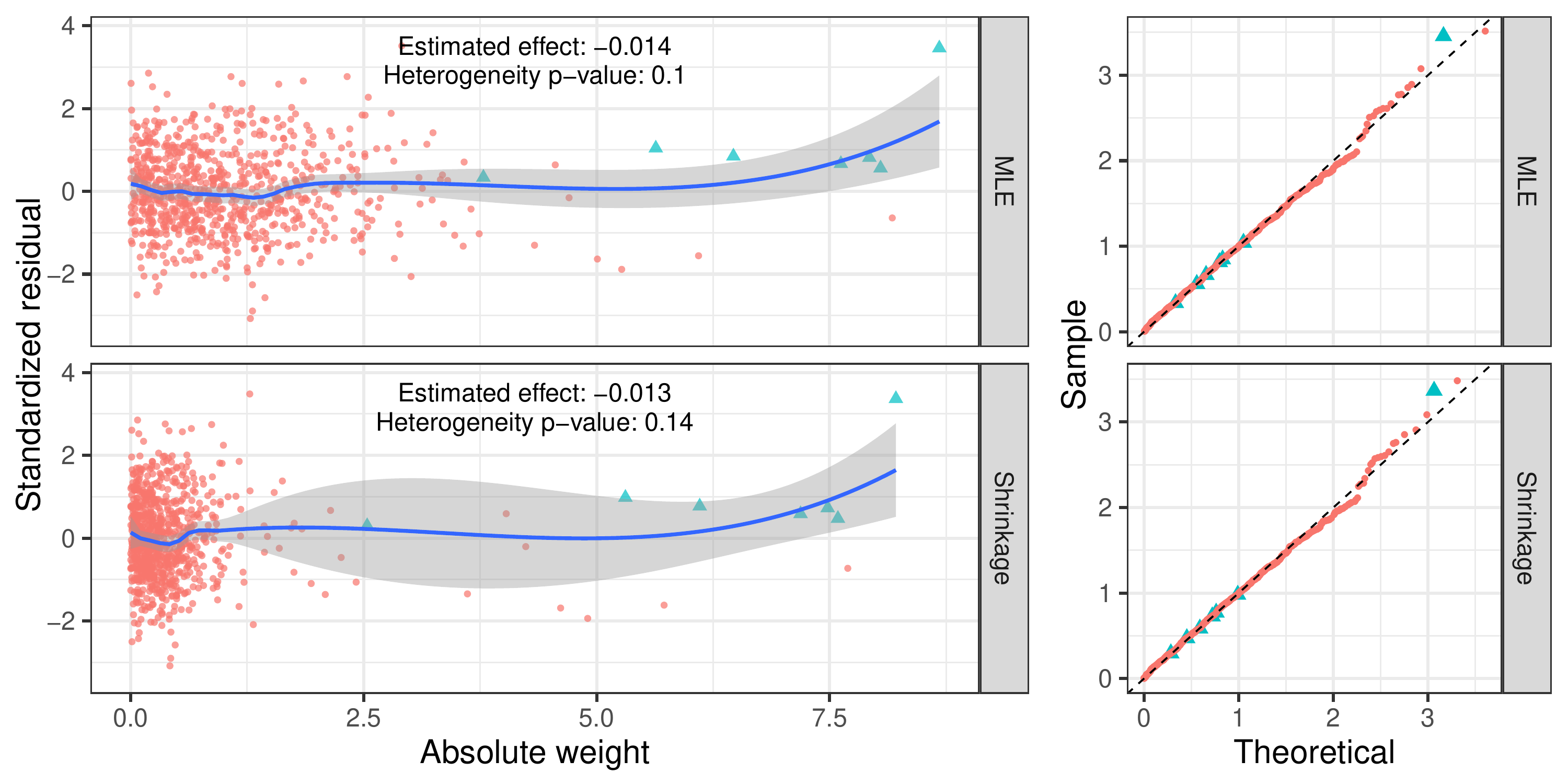}
    \caption{HDL-c.}
    \label{fig:hdc-diagnostics-4}
  \end{subfigure}
  \begin{subfigure}[b]{\textwidth} \centering
    \includegraphics[width = 0.75\textwidth]{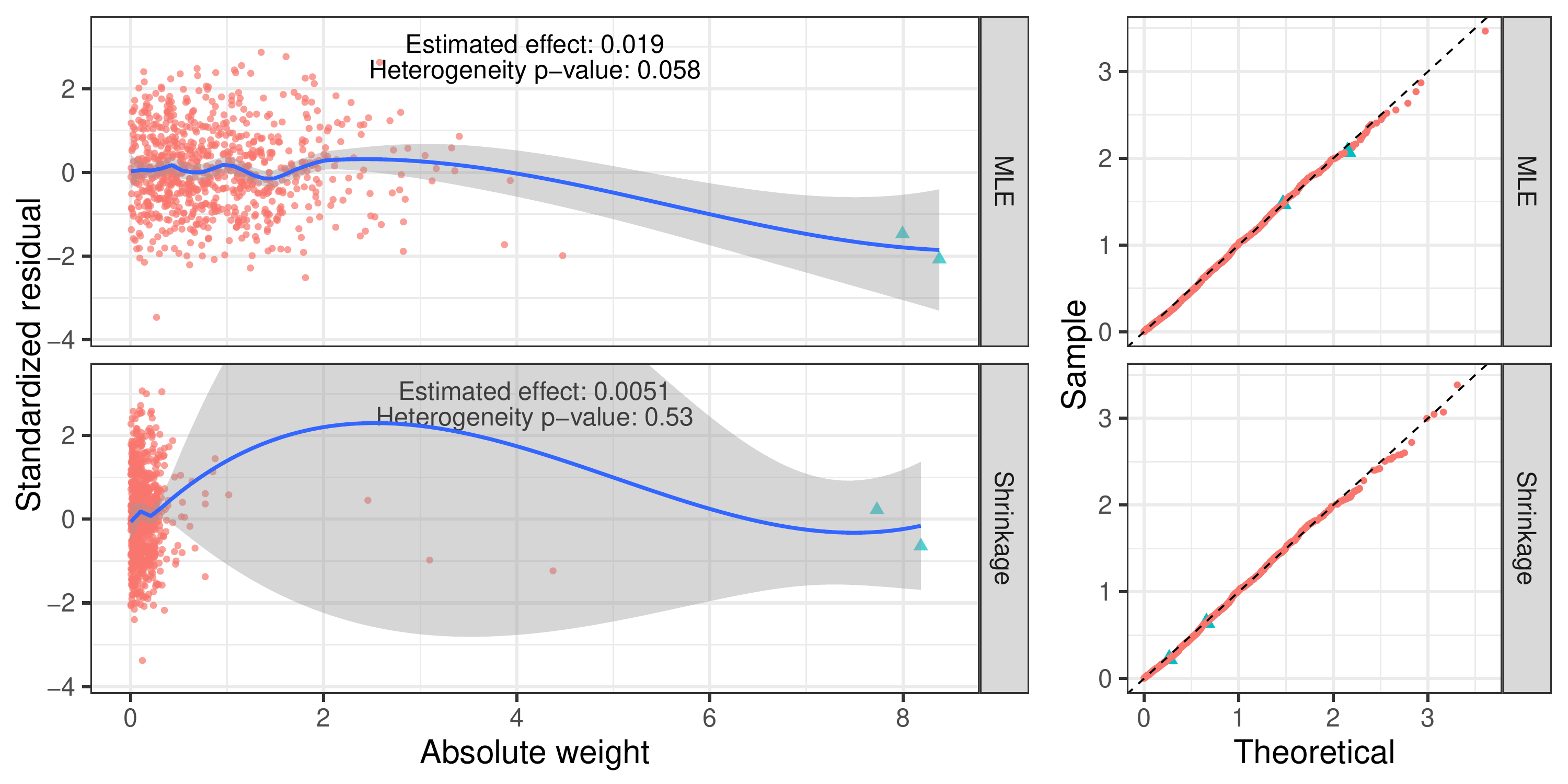}
    \caption{TG.}
  \end{subfigure}
  \caption{Diagnostic plots for the lipids results using restricted
    instruments and the UK BioBank dataset.}
  \label{fig:lipid-diagnostics-4}
\end{figure}

\Cref{fig:bmi-diagnostics,fig:lipid-diagnostics-1,fig:lipid-diagnostics-2,fig:lipid-diagnostics-3,fig:lipid-diagnostics-4}
below are diagnostic plots to check instrument heterogeneity.
If our modeling assumptions are satisfied, most of the standardized
residuals ($y$-axis) should be approximately independent of the
instrument weight ($x$-axis), checked by the scatter-plot in the left
panel of each figure, and should roughly follow the standard normal distribution, checked by
the quantile-quantile plot in the right panel of each figure. The
heterogeneity $p$-values are computed by testing the null model in the
linear regression of the standardized residual on the absolute weight
(expanded using B-splines with degrees of freedom equal to
$\#\text{instruments}/20$).

\end{document}